\documentclass[fleqn,usenatbib]{mnras}
\usepackage{newtxtext,newtxmath}
\usepackage[T1]{fontenc}
\DeclareRobustCommand{\VAN}[3]{#2}
\let\VANthebibliography\thebibliography
\def\thebibliography{\DeclareRobustCommand{\VAN}[3]{##3}\VANthebibliography}
\usepackage{graphicx}
\usepackage{amsmath}
\usepackage{relsize}
\usepackage{booktabs}
\usepackage{orcidlink}

\newcommand{\code}[1]{\texttt{#1}}
\usepackage{color}
\definecolor{purple}{RGB}{107,1,125}

\title[SAMI: Milky Way Analogues]{The SAMI Galaxy Survey: On the importance of applying multiple selection criteria for finding Milky Way Analogues}
\author[S. Tuntipong et al.]{Sujeeporn Tuntipong$^{\orcidlink{0009-0002-8534-5077}}$,$^{1,2}$\thanks{E-mail: stun4076@uni.sydney.edu.au} 
Jesse van de Sande$^{\orcidlink{0000-0003-2552-0021}}$,$^{3,1,2}$ 
Scott M. Croom$^{\orcidlink{0000-0003-2880-9197}}$,$^{1,2}$ 
Stefania Barsanti$^{\orcidlink{0000-0002-9332-5386}}$,$^{2,4}$
\newauthor Joss Bland-Hawthorn$^{\orcidlink{0000-0001-7516-4016}},$$^{1,2}$ 
Sarah Brough$^{\orcidlink{0000-0002-9796-1363}}$,$^{2,3}$ 
Julia J. Bryant$^{\orcidlink{0000-0003-1627-9301}}$,$^{1,2,5}$ 
Sarah Casura$^{\orcidlink{0000-0003-1645-0543}}$,$^{6}$ 
\newauthor Amelia Fraser-McKelvie$^{\orcidlink{0000-0001-9557-5648}}$,$^{7}$ 
Jon S. Lawrence$^{\orcidlink{0000-0002-6998-6993}},$$^{8}$
Andrei Ristea$^{\orcidlink{0000-0003-2723-0810}}$,$^{2,9}$
Sarah M. Sweet$^{\orcidlink{0000-0002-1576-2505}}$ $^{2,10}$ 
\newauthor and Tayyaba Zafar$^{\orcidlink{0000-0003-3935-7018}}$$^{2,11}$ 
\\ \\
$^{1}$Sydney Institute for Astronomy (SIfA), School of Physics, A28, The University of Sydney, NSW, 2006, Australia\\
$^{2}$ARC Centre of Excellence for All Sky Astrophysics in 3 Dimensions (ASTRO 3D), Australia\\
$^{3}$School of Physics, University of New South Wales, NSW 2052, Australia\\
$^{4}$Research School of Astronomy and Astrophysics, The Australian National University, Canberra, ACT 2611, Australia\\
$^{5}$Astralis-USydney, School of Physics, University of Sydney, NSW 2006, Australia\\
$^{6}$Hamburger Sternwarte, Universitaet Hamburg, Gojenbergsweg 112, D-21029 Hamburg, Germany\\
$^{7}$European Southern Observatory (ESO), Karl-Schwarzschild-Strasse 2, 85748 Garching bei Muenchen, Germany\\
$^{8}$Australian Astronomical Optics - Macquarie, Macquarie University, NSW 2109, Australia\\
$^{9}$International Centre for Radio Astronomy Research, The University of Western Australia, 35 Stirling Highway, Crawley, WA 6009, Australia\\
$^{10}$School of Mathematics and Physics, University of Queensland, Brisbane, QLD 4072, Australia\\
$^{11}$School of Mathematical and Physical Sciences, Macquarie University, NSW 2109, Australia\\
}

\date{Accepted 2024 August 20. Received 2024 August 11; in original form 2024 June 21}

\pubyear{\the\year{}}

\begin{document}
\label{firstpage}
\pagerange{\pageref{firstpage}--\pageref{lastpage}}
\maketitle

\begin{abstract}
Milky Way Analogues (MWAs) provide an alternative insight into the various pathways that lead to the formation of disk galaxies with similar properties to the Milky Way. In this study, we explore different selection techniques for identifying MWAs in the SAMI Galaxy Survey. We utilise a nearest neighbours method to define MWAs using four selection parameters including stellar mass ($M_{\star}$), star formation rate ($SFR$), bulge-to-total ratio ($B/T$) and disk effective radius ($R_{\rm{e}}$).  Based on 15 different selection combinations, we find that including $M_{\star}$ and SFR is essential for minimising biases in the average MWA properties as compared to the Milky Way. Furthermore, given the Milky Way's smaller-than-average size, selection combinations without $R_{\rm{e}}$ result in MWAs being too large. Lastly, we find that $B/T$ is the least important parameter out of the four tested parameters. Using all four selection criteria, we define the top 10 most Milky Way-like galaxies in the GAMA and Cluster regions of the SAMI survey. These most Milky-Way-like galaxies are typically barred spirals, with kinematically cold rotating disks and reside in a wide range of environments. Surprisingly, we find no significant differences between the MWAs selected from the GAMA and Cluster regions. Our work highlights the importance of using multiple selection criteria for finding MWAs and also demonstrates potential biases in previous MWA studies.
\end{abstract}



\section{Introduction}

As our home, the Milky Way is the galaxy that we are most familiar with and consequently is a benchmark for understanding disk galaxies \citep{BH2016}.  Our position within the Milky Way is both an advantage and a disadvantage. Several galactic surveys, such as the Apache Point Observatory Galactic Evolution Experiment \citep[APOGEE;][]{AllendePrieto2008}, Gaia \citep{Gaia2021}, and the GALactic Archaeology with HERMES survey \citep[GALAH;][]{Buder2021} provide an exquisite view of the Milky Way structure.  However, our sight lines towards the centre of the Galaxy are heavily obscured by dust and hence we do not have a complete picture of what the Milky Way would look like observed from an extragalactic perspective.  To get a different perspective on how galaxies like the Milky Way might have formed and evolved, studies of Milky Way-like galaxies, often called Milky Way Analogues (MWAs), have been proposed. The use of MWAs is based on the Copernican assumption that the Milky Way is not unique \citep{LNB2015,FraserMcKelvie2019} and therefore its properties should be relatively comparable to other disk galaxies.

Identifying MWAs is typically based on comparing the Milky Way to other galaxies using a set of parameters such as stellar mass ($M_{\star}$), star formation rate (SFR), morphology, or size \citep[e.g.][]{Boardman2020b}. The most commonly used parameter is the total galaxy stellar mass $M_{\star}$, which has been applied in all observational studies focussed on MWAs because most other galaxy parameters strongly correlate with $M_{\star}$ \citep[e.g.][]{Baldry2006,Robotham2013}.

Star formation rate (SFR) is the second most important parameter and is strongly correlated with stellar population and structural properties, as well as environment \citep[e.g.][]{Petropoulou2011,Owers2019,Steffen2021,FraserMcKelvie2022}.  SFR is therefore a natural property for selecting MWAs \citep{Boardman2020b}. Colour, which serves as a proxy for SFR (e.g. from H$\alpha$ or UV, FIR) is also commonly used, as there is a strong (although not one-to-one) relationship between SFR and colour \citep{Salim2015} and colours are directly observable in large observational imaging surveys \citep{Krishnarao2020b,Scott2018}.

The Milky Way is morphologically classified as a late-type, barred, spiral galaxy (SBc) \citep[e.g.][]{Englmaier1999,Churchwell2009,Pettitt2014,Pettitt2015}, with a low bulge-to-total ratio ($B/T$) due to the small bulge component and the dominating disk component. Therefore, morphology is a third important parameter for identifying MWAs as the presence of a bar and/or spiral feature is essential for understanding the properties of galaxies \citep{Krishnarao2020b}. Instead of relying on visual classification, another method is to use $B/T$ for the selection of MWAs, where the $B/T$ is derived from fitting two S\'ersic profiles (e.g. a bulge and a disk) to a galaxy's surface brightness. However, $B/T$ measurements can be uncertain as the bar component is often ignored or poorly fit by the central S\'ersic profile \citep{Lange2016}. Another structural parameter used to select MWAs is the size, commonly defined as the disk scale length $R_{\rm{d}}$ or the disk effective radius $R_{\rm{e}}$. However, several studies have shown that the Milky Way's size is atypical, where at fixed $M_{\star}$ the Milky Way is a factor of two smaller than the average disk galaxy \citep{LN2016,Boardman2020b}. 

The choice of which set of selection parameters to use has depended on the aims of studies \citep{Boardman2020b}. For example, in order to study the progenitors of the MWA \citet{vanDokkum2013} and \citet{Santistevan2020} used only halo mass $M_{\rm{halo}}$ or stellar mass $M_{\star}$, because these two studies specifically aimed for growth of the Milky Way as a function of time in simulations. On the other hand, \citet{Krishnarao2020b} used three parameters including $M_{\star}$, morphology and $(NUV-r)$ colour to investigate the effects of the bar component on the ionisation mechanisms of gas in galaxies. 

The advent of large integral field spectroscopic (IFS) surveys (e.g. the Calar Alto Legacy Integral Field Area Survey \citep[CALIFA;][$\sim$600 galaxies]{Sanchez2012}, the Sydney-AAO (Australian Astronomical Observatory) Multi-object IFS survey \citep[SAMI;][3068 galaxies]{Croom2012}, and the Mapping Nearby Galaxies at Apache Point Observatory \citep[MaNGA;][$\sim$10000 galaxies]{Bundy2015}), creates an opportunity to study MWA in more detail. The large-scale IFS surveys can reveal the detailed internal structure of galaxies \citep{Croom2021}, including stellar kinematics and stellar populations resolved at a $\sim1$ kpc resolution. These spatially-resolved spectroscopic measurements are fundamental for understanding many aspects of galaxy evolution such as quenching, gas flows and environmental effects. 

In this work, we selected MWAs through a nearest neighbours method using galaxies observed by the SAMI survey \citep{Croom2012}. We are particularly interested in how different selection criteria impact the types of MWAs that are found. Our MWA selection is based on four selection criteria including total stellar mass ($M_{\star}$), star formation rate ($SFR$), bulge-to-total ratio ($B/T$) and effective radius of the disk ($R_{\rm{e}}$). The IFS data from the SAMI survey provides an opportunity to explore the kinematic properties of the selected MWAs based on various combinations of selection parameters as well as different environments, from low-density regions to groups and clusters. 

Section \ref{sec:data and methods} introduces the data, flags applied to the data, selection methods, and selection parameters. Section \ref{sec:diff_selection} presents the  MWAs based on different combinations of selection parameters. In Section \ref{sec:MWAs_prop} we describe the properties of the top-10 MWAs including morphology, stellar kinematics, star formation, $(g-i)$ colour, and environments. Section \ref{sec:discussion} explains how different selection methods and strategies affect the properties of MWAs, and includes a discussion where we compare our results to those from previous studies. Finally, all outcomes are summarised in Section \ref{sec:conclusions}. Throughout the paper we assume a $\Lambda$CDM cosmology with $\Omega_\mathrm{m}$=0.3, $\Omega_{\Lambda}=0.7$, and $H_{0}=70$ km s$^{-1}$ Mpc$^{-1}$.

\section{Data and Methods}
\label{sec:data and methods}

\subsection{SAMI data}

The SAMI Galaxy survey is a large, multi-object IFS survey \citep{Croom2012, Bryant2015} that aims to understand how environment and gas flows influence galaxy evolution, how mass and angular momentum evolve within galaxies, and compare its observational results to large-volume cosmological simulations. The SAMI Galaxy survey observations were carried out between 2013 and 2018, which resulted in $3068$ unique galaxies being observed across a wide range of mass and environments with spatially resolved structural and kinematics measurements \citep{Croom2021}.

The target selection for the SAMI survey was categorised into two groups. The first was the selection of GAMA galaxies \citep{Bryant2015} within three $4^{\circ}\times12^{\circ}$ observed regions of the GAMA survey \citep{Driver2011}. Another was the Cluster galaxies \citep{Owers2017} from eight regions overlapping either the Sloan Digital Sky Survey \citep[SDSS;][]{Ahn2012} or the VLT Survey Telescope ATLAS Survey \citep[VST;][]{Shanks2015}. The targets were situated within the stellar mass range $10^{7}<M_{\star}/\rm{M}_{\odot}<10^{12}$ and redshift $z<0.115$, which gave four volume-limited samples. The complete SAMI data set was released in Data Release 3 \citep{Croom2021}.

The stellar kinematics of the SAMI galaxies were derived by extracting the line-of-sight stellar velocity distribution \citep[LOSVD;][]{vandeSande2017b} using the penalised pixel fitting code \citep[\textsc{pPXF};][]{Cappellari2004, Cappellari2017}. One of the key stellar kinematic data products was the spin parameter $\lambda_{R}$, which quantifies the ratio of ordered-to-random motion of a galaxy within one effective radius. For the SAMI survey, $\lambda_{R}$ was calculated in \citet{vandeSande2017b} and defined according to \citet{Emsellem2007} as:
\begin{equation}
    \lambda_{R}=\frac{\langle R|V \rangle}{\langle R \sqrt{V^{2}+\sigma^{2}} \rangle}=\frac{\sum^{N_{\rm{spx}}}_{n=0}F_{n}R_{n}|V_{n}|}{\sum^{N_{\rm{spx}}}_{n=0}F_{n}R_{n}\sqrt{V^{2}+\sigma^{2}}},
    \label{eq:spin_parameter}
\end{equation}
where the subscript $n$ refers to a spaxel position within an elliptical aperture, $N_{\rm{spx}}$ is the total number of spaxels, $V_{n}$ is the stellar velocity in km s$^{-1}$, $\sigma_{n}$ is the velocity dispersion in km s$^{-1}$, $F_{n}$ is the flux of the $i$th spaxel in units of erg cm$^{-2}$s$^{-1}$\AA$^{-1}$ and $R_{n}$ is the radius in arcseconds. The radius used the semi-major axis of an elliptical aperture measured by a multi-Gaussian Expansion (MGE) fit.

The spin parameter $\lambda_{R}$ was converted into the edge-on spin parameter $\lambda_{R,\rm{EO}}$ which is the maximum $\lambda_{R}$ measured when a galaxy is edge-on oriented to an observer. The edge-on spin parameter $\lambda_{R,\rm{EO}}$ and the measured spin parameter $\lambda_{R}$ are related by
\begin{equation}
    \lambda_{R}=C\left(i\right)\frac{\lambda_{R}}{\sqrt{1+\lambda^{2}_{R,\rm{EO}}\left(C^{2}\left(i\right)-1\right)}},
\end{equation}
\label{eq:spin_parameter1}
where 
\begin{equation}
    C\left(i\right)=\frac{\sin{\left(i\right)}}{\sqrt{1-\beta\cos^{2}{\left(i\right)}}},
\end{equation}
\label{eq:spin_parameter2}
where $i$ is the observed inclination and $\beta$ is the anisotropy under an oblate assumption \citep{Binney2005,Cappellari2007,Emsellem2007,Harborne2018}. 

The ellipticity $\varepsilon$ for SAMI is defined as the averaged ellipticity within one effective radius measured from the best-fitting MGE model \citep{vandeSande2017b}. The intrinsic ellipticity $\varepsilon_{i}$ is defined as the ellipticity when the galaxy is edge-on oriented, which is obtained by tracking its location along the $\varepsilon_{i}$ line on the $\lambda_{R}-\varepsilon$ galaxy model \citep{Cappellari2007,Emsellem2011,Cappellari2016,vandeSande2017b,vandeSande2018}. Note that only galaxies above the edge-on line in the $\lambda_{R}-\varepsilon$ plane are corrected for edge-on spin parameter $\lambda_{R,\rm{EO}}$ and the intrinsic ellipticity $\varepsilon_{i}$, which allows the comparison of the stellar kinematics independent of the observed inclination. 

All SAMI galaxies were visually classified and inspected for bars and signs of interaction. The classification was done by 3 or 4 team members, except 305 galaxies in APMCC 917, Abell 4038, EDCC 442, and Abell 3880 that were classified by AFM. The visual classification was performed on colour images from three surveys: Hyper Suprime-Cam \citep[HSC;][]{Miyazaki2018}, the Dark Energy Camera Legacy Survey \citep[DECaLS;][]{Dey2019} or the Panoramic Survey Telescope and Rapid Response System \citep[Pan-STARRS;][]{Flewelling2020}. The best image quality was from HSC (FWHM$_{\rm{PSF},i}$~=~0\farcs67), followed by DeCaLS (FWHM$_{\rm{PSF},i}$~=~1\farcs11), and Pan-STARRS (FWHM$_{\rm{PSF},i}$~=~1\farcs30). We note that HSC imaging was not available for all galaxies. We required a minimum classification fraction of $1/3$ to indicate the presence of a bar or signs of interactions. The visual classification catalogue will be made available on Vizier.

We obtained star formation rates of SAMI galaxies from \citet{Ristea2022} where three different SFRs estimates were combined to create the largest sample. The first set of SFRs was derived by the fitting spectral energy distribution (SED) of a galaxy with Code Investigating GALaxy Emission \citep[CIGALE;][1891 galaxies]{Burgarella2005,Noll2009,Boquien2019} from the Galaxy Evolution Explorer (GALAX) SDSS Wide-field Infrared Survey Explorer (WISE) Legacy Catalogue version 2 \citep[GSWLC-2;][]{Salim2016,Salim2018}. The second set of SFRs was obtained using the SED fitting code PROSPECT \citep[][657 galaxies]{Robotham2020}. The last set came from a combination of NUV photometry and WISE W3 fluxes \citep[][81 galaxies]{Janowiecki2017}. These SFRs were re-scaled to be consistent with the cosmology of the SAMI Survey. Using this combination of the three different datasets, there are 2629 SAMI galaxies with the SFRs available which accounted for 86\% of the SAMI sample.

Luminosity-weighted, single stellar population equivalent stellar ages, metallicities $[\rm{Z}/\rm{H}]$ and $\alpha-$abundance $[\alpha/\rm{Fe}]$ of SAMI galaxies were obtained from \citet{Scott2017}. These measurements were determined by integrating spectra and deriving Lick absorption line strength indices and stellar population parameters within a series of standard, fixed apertures. In our study, we use the measurements within 1 $R_{\rm e}$.

\subsection{Parent sample selection}
\label{subsec:flagging}

The SAMI galaxy survey provides a unique opportunity to investigate the possibility of locating MW-like galaxies near or within dense large-scale structures such as massive groups or clusters.
To distinguish between the most massive cluster 
($\log(M_{\rm{halo}}/\rm{M_{\odot}})>14.25 $) and galaxies in groups, pairs or isolation, we separate the data according to SAMI source fields, i.e. the GAMA and Cluster regions which reflect this environmental difference.
Although we expect that MWAs selected in clusters may have some biases (e.g. tendency to early types and lower SFRs), we used both GAMA and Cluster samples to investigate the impact of all environments on the MWA selection.

\begin{figure*}
    \centering
    \includegraphics[scale=0.7]{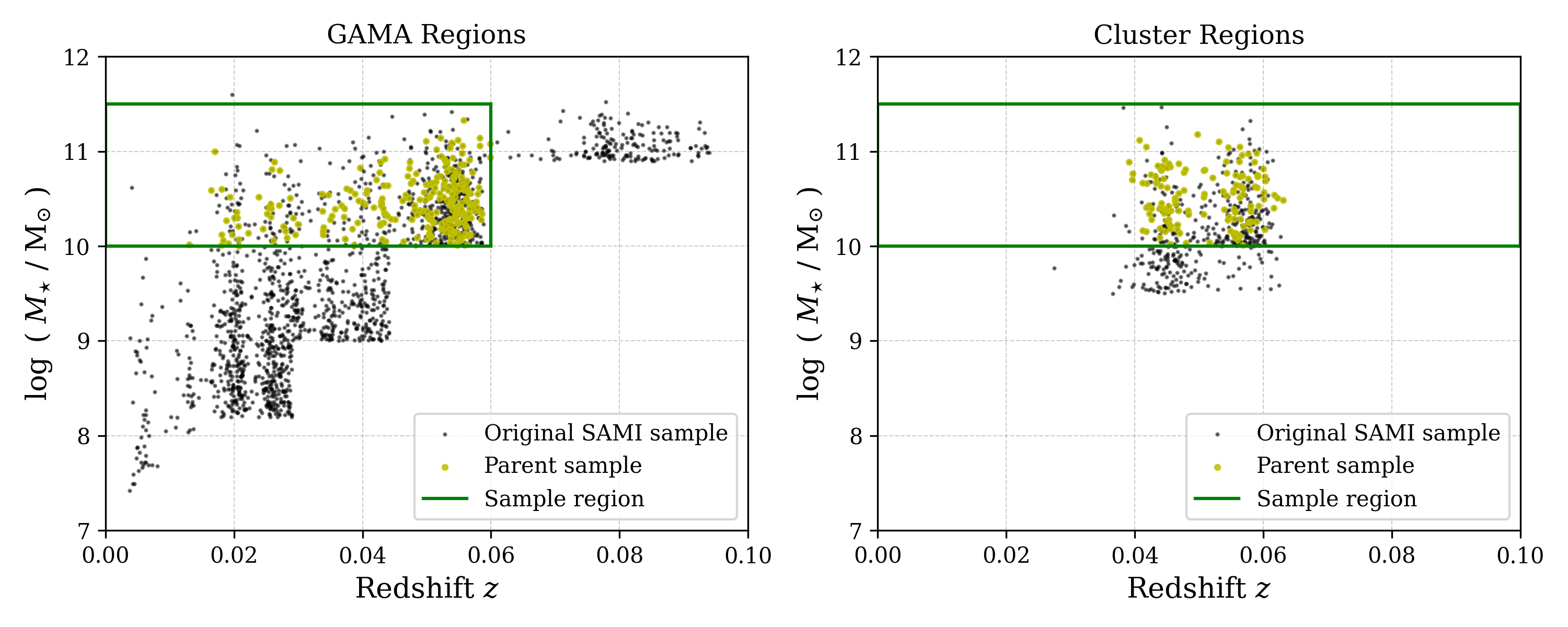}
    \caption{The final samples (green dots) in the volume-limited region (green edges) selected from the SAMI parent population (black dots) are shown separately for the GAMA (left panel) and Cluster (right panel) regions. The stellar mass selection was $10.0<\log(M_{\star}/\rm{M}_{\odot})<11.5$. While the redshift was limited to $z<0.06$. Note that some Cluster galaxies marginally exceed this cut due to peculiar motions.}
    \label{fig:z_M}
\end{figure*}

Before applying the MWA selection, we flagged any spurious data to create a cleaner sample of galaxies. Table \ref{tab:data} lists all catalogues containing the key information of the galaxies that were used for flagging. We applied a stellar mass cut of $10<\log(M_{\star}/\rm{M}_{\odot})<11.5$ for the GAMA and Cluster sample, as well as a redshift cut of $z<0.06$. Note that some of the Cluster galaxies exceeded the $z<0.06$ limit due to the peculiar motions in the clusters \citep{Owers2017}, but for completeness we decided not to remove these. The mass and redshift cut brings the sample to 1412 galaxies with 790 and 622 galaxies in the GAMA and Cluster regions, respectively. The stellar mass and redshift limits are illustrated in Fig. \ref{fig:z_M}. 

For both the GAMA and Cluster samples, we selected only galaxies with reliable $r-$band double-component S\'ersic surface brightness profile fits. In these fits, the S\'ersic index $n$ for the central component was left as a free parameter, whereas an exponential ($n=1$) profile was used for the disk. We only include galaxies with the double-component S\'ersic fits as our method considers $B/T$ as a selection parameter. In principle, we could include galaxies where a single component disk fit was preferred.  However, testing this we find it makes no difference and none are included in our MWA sample as they are $\sim3\sigma$ away from the Milky Way's $B/T$. As the double-component S\'ersic fits in the GAMA and Cluster samples were derived from different imaging campaigns, we applied different flags to both samples.

For the GAMA sample, we used the \code{BDModelsv03} from \citet{Casura2022}, which is based on photometry from the Kilo-Degree Survey \citep[KiDS;][]{Kuijken2019}. First, we only selected objects where the double component $r-$band  fit was preferred over a single S\'ersic component (\code{R\_NCOMP} $=2.0$). Secondly, the S\'ersic-profile central effective radius had to be larger than the pixel size (\code{R\_D\_B\_RE} $> 0.2$) to ensure that the central component is resolved. Thirdly, we kept galaxies if the $r-$band point spread functions (PSF) measurements were available, and if the centres of the galaxies were not masked (\code{R\_D\_BDQUAL\_FLAG} equal to 0, 32, 64 or 96). Fourthly, we excluded galaxies where the central component S\'ersic index was smaller than 0.1 or larger than 15 ($0.1<$\code{R\_D\_B\_NSER} $< 15$). Lastly, we only selected galaxies with effective disk radii at least $5\%$ greater than the effective radii of the central component (\code{R\_D\_D\_RE} $> 1.05$ $\times$ \code{R\_D\_B\_RE}) to ensure that central and disk profiles represent the inner and outer regions of the galaxy, respectively. After applying these cuts, 285 GAMA galaxies remain.

We selected the Cluster galaxies following five criteria defined by \citet{Barsanti2021b}. First, we selected galaxies where the differences in logarithmic marginal likelihoods (LML) between the central and disk components were greater than 60, as these galaxies were likely to be double-component galaxies (\code{r-LBF} $> 60$). Secondly, we selected galaxies with S\'ersic profile indices of the central component between 0.1 and 15 ($0.1 <$ \code{r-d-nser1} $<15$). Thirdly, we kept the galaxies if the effective radii of both components were not pegged to either upper or lower limits (\code{r-d-npegged} $=0$ and \code{r-d-npegged-warn $=0$}). Fourthly, we kept the galaxies if the effective radii of the disks were greater than $1.05$ times of the effective radii of the bulges (\code{r-d-re2} $> 1.05$ $\times$ \code{r-d-re1}). Lastly, we kept the galaxies if the effective radii of the bulges were greater than $80\%$ of the Point Spread Function (PSF) full-width at half maximum (FWHM) (\code{r-d-re1} $> 0.8$ $\times$ \code{r\_psf\_fwhm}). After applying these cuts, 274 Cluster galaxies remain. 
We note that most of the SAMI galaxies on both GAMA and Cluster regions were removed from the parent sample selection based on the single-component fit being a better representation of the surface brightness profile than the double-component fit. After the outlier removal, a total of 559 galaxies remain in the sample.

\begin{table*}
\small
\centering
 \caption{A summary of the imported catalogues providing the data for selection of MWAs.}
 \label{tab:data}
 \begin{tabular}{lll}
  \hline
  Name & Description & Reference\\
  \hline
  InputCatGAMADR3 & Target and properties in the GAMA regions & \citet{Bryant2015}\\
  InputCatClusterDR3 & Target and properties in the Cluster regions & \citet{Owers2017}\\
  CubeObs & Observed cubes and quality flags & \citet{Croom2021}\\
  BDModelsv03 & All-band Photometric models in the GAMA regions & \citet{Casura2022}\\
  SDSS\_ATLAS\_rband & $r-$band Photometric models in the Cluster regions & \citet{Barsanti2021b}\\
  VisualMorphologyDR3 & Visual morphological classification & \citet{Cortese2016}\\
  SSPAperturesDR3 & Simple Stellar Population age, metallicity and $\alpha-$abundance & \citet{Scott2017} \\ 
  IndexAperturesDR3 & Stellar continuum index measurements & \citet{Scott2017} \\
  DensityCatDR3 & Local density estimation & \citet{Croom2021,Brough2017} \\
  samiDR3Stelkin & Stellar kinematics measurements & \citet{vandeSande2017b} \\
  SAMI\_DR3\_SFR\_rescaled & Re-scaled SFRs based on SED fitting or NUV-WISE magnitudes   & \citet{Ristea2022} \\
  SAMI\_detailed\_classifications& Visual classification for bars and signs of interaction & This paper\\
  \hline
 \end{tabular}
\end{table*}

We also introduced a flag for the Lick absorption line strength indices measurements. We used the elliptical aperture with a semi-major axis radius measured by an MGE fit (\code{FLAG\_RE\_MGE}) similar to the aperture utilised for the stellar kinematic measurements. Only one galaxy from the GAMA regions was flagged.  Furthermore, we did not use galaxies without SFR measurements in the \code{SAMI\_DR3\_SFR\_rescaled} catalogue or without stellar kinematic measurements in the \code{samiDR3Stelkin} catalogue. Lastly, we flagged any remaining galaxies if the measurements of the disk radius were contaminated by nearby sources, leading to unphysically large disk effective radii ($>20$ kpc). With all flags incorporated, the final sample contained 340 galaxies, with 254 and 134 in the GAMA and Cluster regions, respectively.

\subsection{Method}
\label{sec:method}

We selected MWAs separately in the GAMA and Cluster regions but using the exact same method. The MWAs were selected using the nearest neighbours method. This method assumes that all measurement errors are independent and follow a Gaussian distribution. We calculated a distance, $D$, from the Milky Way to other galaxies in the multi-dimensional parameter space of our selection parameters. In each dimension, the distance was normalised by the uncertainty on the Milky Way value of that parameter, such that $D$ is given by
\begin{equation}
    D=\sqrt{~\sum^{n}_{i=0}~~\left(\frac{x_{i}-x_{i,\textrm{MW}}}{\sigma_{i,\textrm{MW}}}\right)^{2}},
	\label{eq:nearest_neighbours}
\end{equation}
where $x_{i}$ are the parameters of the galaxy, the subscript MW marks the same parameters for the Milky Way and $\sigma_{i,\textrm{MW}}$ indicates the uncertainties for the Milky Way values. By sorting galaxies based on their distance $D$ to the Milky Way, we selected galaxies most likely to be analogues of the Milky Way. Note that we did not include the uncertainties for the SAMI galaxies, as this may bias us in selecting objects with large uncertainties.

\subsection{Selection parameters}

The parameters that we used for the selection of Milky Way Analogues were stellar mass ($M_{\star}$), star formation rate ($SFR$), $r-$band bulge-to-total ratio ($B/T$) and $r-$band disk effective radius ($R_{\rm{e}}$). Table \ref{tab:ref} summarises the reference parameters applied for the selection. Depending on the dynamic range of the parameter, we either used a logarithmic scale ($\log_{10} M_{\star}$, $\log_{10}SFR$), or a linear scale ($B/T$, $R_{\rm{e}}$).

\subsubsection{Stellar mass ($M_{\star}$)}

The stellar masses $M_{\star}$ of the SAMI galaxies were calculated using the observed-frame $(g-i)$ colour, $i-$band magnitude and redshift \citep{Bryant2015}, based on the relation derived from SED fitting of GAMA galaxies \citep{Taylor2011}. The Milky Way $M_{\star}$ was estimated by \citet{BH2016} to be $(5\pm1)\times10^{10} ~\rm{M}_{\odot}$, or $10.70\pm0.09$ on a logarithmic scale. This is a summation of the mass of bulge and disk components derived from a dynamical model fitting to stellar surveys \citep{Reid2014,LopezCorredoira2014,Kafle2012} and to the Galactic rotational curve \citep{Sofue2009,Koposov2010,Kupper2015}.

\subsubsection{Star formation rate ($SFR$)}

The SFRs of the SAMI galaxies were derived either from SED fitting with CIGALE, SED fitting with PROSPECT, or NUV-WISE magnitudes \citep{Ristea2022}. A re-scaling scheme was applied to the SFRs so that the values were consistent with the SAMI cosmology. The SFR used for the Milky Way is $1.78\pm0.36~\rm{M}_{\odot}yr^{-1}$, or $0.25\pm0.09$ on a logarithmic scale. We derived this number by combining 12 previous studies \citep{Chomiuk2011,Elia2022}, normalised to a Kroupa IMF \citep{KroupaWeidner2003,Kennicutt2009}. Note that the conversion from a Kroupa to Chabrier IMFs is small, hence we did not apply a correction between the IMFs used for the SAMI data and the Milky Way. The details of the MW SFR calculation are presented in Appendix \ref{appendix: SFR}.

\subsubsection{Bulge-to-total ratio ($B/T$)}

The bulge-to-total ratios $B/T$ of the SAMI galaxies in the $r-$band were derived from photometric fitting \citep{Barsanti2021b,Casura2022} using the \textsc{PROFIT} code \citep{Robotham2017}. The two components were fitted using an exponential profile for the disk and a S\'ersic profile for the central component. Note that the bars were not separately considered in this fitting process, so flux from a bar could contribute to either component.

We estimated the Milky Way's $B/T$ from the ratio between the bulge stellar mass and the total stellar mass of the Milky Way. Similar to the SFR, in Appendix \ref{appendix: B/T} we re-derived the bulge stellar mass to be $(1.49\pm0.43)\times10^{10}~\rm{M}_{\odot}$ but adopt the total stellar mass of $(5\pm1)\times10^{10}~\rm{M}_{\odot}$ from \citet{BH2016}. We obtained the Milky Way $B/T$ at $0.30\pm0.10$, consistent with previous studies \citep{LN2015,BH2016,Barbuy2018}. Note that this $B/T$ includes the mass of the Milky Way's bar structure.

\subsubsection{Disk effective radius ($R_{\rm{e}}$)}

In the SAMI survey, the $r-$band disk effective radius of the disk component was computed from photometric fitting \citep{Barsanti2021b,Casura2022}. The intensity profile of the disk component was assumed to be exponential. 
The Milky Way disk scale length in the $r-$band was estimated to be $2.3\pm0.6$ kpc by \citet{Hammer2007,Yin2009}. The disk scale length was derived by averaging scale lengths from 14 data sets measured using different approaches. The uncertainty was derived by averaging the errors associated with the scale lengths and then doubling the value to account for possible systematic errors \citep{Hammer2007}. 
We converted the exponential disk scale length to an effective radius $R_{\rm{e}}$ using
\begin{equation}
    R_{\rm{e}}=1.678R_{\rm{d}},
	\label{eq:scale_length_to_effective_radius}
\end{equation}
where $R_{\rm{d}}$ is the disk scale length \citep{Graham2005}. This resulted in an $r-$band disk effective radius of $3.86\pm1.01$ kpc for the Milky Way.

\begin{table*}
\centering
 \caption{A summary of the Milky Way's selection parameters providing the value with uncertainty, the logarithmic value with uncertainty (if available) and the reference.}
 \label{tab:ref}
 \begin{tabular}{lccl}
  \hline
  Parameter & Value & Logarithmic value & Reference\\
  \hline
   Stellar mass $M_{\star}$ ($\rm{M}_{\odot}$) & $(5\pm1)\times10^{10}$  & $10.70\pm0.09$ & \citet{BH2016}\\
   Star formation rate $SFR$ ($\rm{M}_{\odot}\rm{yr}^{-1}$) & $1.78\pm0.36$  & $0.25\pm0.09$ & Derived in this paper\\
   Bulge-to-total ratio $B/T$ & $0.30\pm0.10$ & - &  Derived in this paper\\
   Disk Effective radius $R_{\rm{e}}$ (kpc) & $3.86\pm1.01$ & $0.59\pm0.11$ &  \citet{Hammer2007,Yin2009}\\
  \hline
 \end{tabular}
\end{table*}

\section{The impact of different MWA selection parameters}
\label{sec:diff_selection}

In this section, we illustrate the impact of using different selection combinations based on the four key galaxy parameters. The main focus of this section is on the results using the sub-sample of galaxies from GAMA regions. However, we do discuss the Cluster selection where appropriate.

\begin{figure*}
    \centering
    \small
    \includegraphics[scale=0.208]{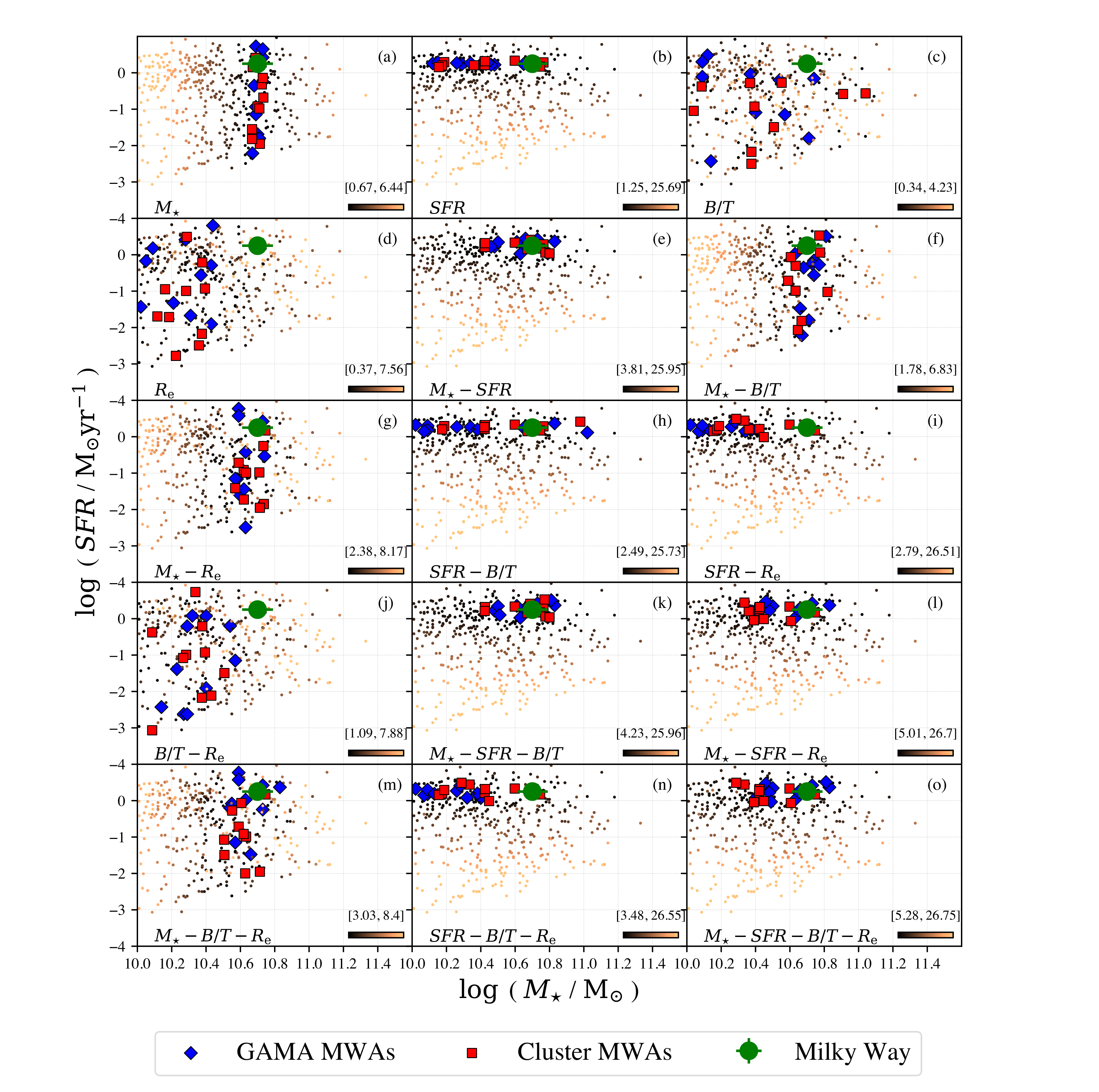}
    \caption{$\log(M_{\star})$ versus $\log(SFR)$ with the top-10 MWAs from the GAMA (blue diamonds) and Cluster (red squares) regions. The green dot with $1\sigma$ error bars shows the location of the Milky Way. The colour coding of the small data points represents the distance (Eq. \ref{eq:nearest_neighbours}) of the sample from the Milky Way value, with the scale given by the colour bar in the lower right corner of each diagram. The resulting MWAs from 15 different parameter selection combinations (as displayed in the bottom left corner of each diagram) are distributed differently in the $\log(M_{\star})-\log(SFR)$ plane.}
    \label{fig:M_SFR_all}
\end{figure*}

\begin{figure*}
    \centering
    \small
    \includegraphics[scale=0.208]{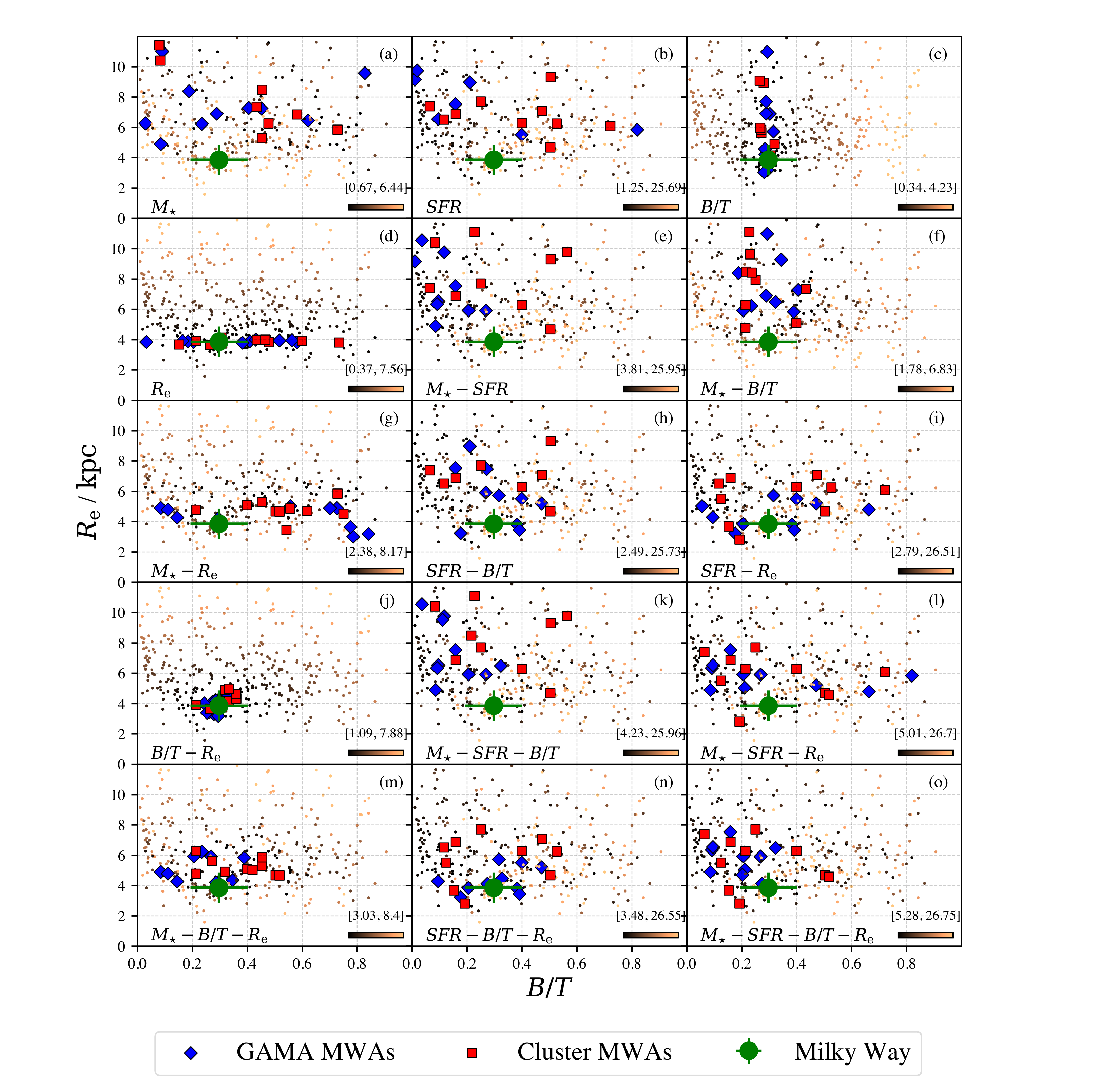}
    \caption{$B/T$ versus $R_{\rm{e}}$ with the top-10 MWAs from the GAMA (blue diamonds) and Cluster (red squares) regions. The green dot with $1\sigma$ error bars shows the location of the Milky Way. The colour coding of the small data points represents the distance (Eq. \ref{eq:nearest_neighbours}) of the sample from the Milky Way value, with the scale given by the colour bar in the lower right corner of each diagram. The figures illustrate that the resulting MWAs from 15 different parameter selection combinations (as displayed in the bottom left corner of each diagram) are distributed differently in the $B/T-R_{\rm{e}}$ plane.}
    \label{fig:BT_RE_all}
\end{figure*}

To investigate the impact of different selection parameter combinations for identifying MWAs, we begin by focusing on a subset of galaxies. We select 10 galaxies with the smallest distance $D$ based on Eq. \ref{eq:nearest_neighbours}, which account for 4\% and 7\% of our GAMA and Cluster samples used in this analysis. The number of galaxies in the subset is picked to be large enough to make a statistical comparison, but small enough to avoid the averages of subsamples getting too close to the median of the entire samples. We discuss the impact of varying the number of galaxies being selected as MWAs in Appendix \ref{appendix:bias}.

Fig. \ref{fig:M_SFR_all} illustrates the resulting MWAs selected by different parameter combinations on the $\log(M_{\star})-\log(SFR)$ plane. Fig. \ref{fig:M_SFR_all}a-d show the impact of using only one parameter to select MWAs. When we use $\log(M_{\star})$, we find that the typical MWAs have lower SFRs than the Milky Way (Fig. \ref{fig:M_SFR_all}a). But when we use $\log(SFR)$, $B/T$, or $R_{\rm{e}}$ alone (Fig. \ref{fig:M_SFR_all}b-d) we find that the MWAs are biased towards low stellar masses. Fig. \ref{fig:M_SFR_all}e-o show the impact of using more than one parameter to select MWAs. When using a combination of $\log(M_{\star})$ and $\log(SFR)$ (Fig. \ref{fig:M_SFR_all}e), the MWAs distribute closely and symmetrically around the value of the Milky Way, as is expected given the selection. The MWAs spread out more spatially when adding $B/T$ and/or $R_{\rm{e}}$ into the $\log(M_{\star})-\log(SFR)$ selection combination. Overall, when using a selection combination with $\log(M_{\star})$ included, the MWAs distribute loosely around the Milky Way's value. On the other hand, excluding $\log(M_{\star})$ from a selection combination provides MWAs biased towards lower stellar masses. Likewise, there is a bias towards passive MWAs with low SFRs when we exclude the $\log(SFR)$ from selection combinations. 

Similarly, Fig. \ref{fig:BT_RE_all} illustrates the resulting MWAs selected by different parameter combinations on the $B/T-R_{\rm{e}}$ plane. The MWAs are biased towards larger sizes when we exclude $R_{\rm{e}}$ from selection combinations. On the other hand, the $B/T$ of the MWAs are distributed symmetrically around the Milky Way value. The MWAs are distributed closely and symmetrically around the Milky Way's value when using a combination of $B/T-R_{\rm{e}}$ (Fig. \ref{fig:BT_RE_all}j). The MWAs spread out more spatially when adding $\log(M_{\star})$ and/or $\log(SFR)$ into the $B/T-R_{\rm{e}}$ selection combination.

\begin{figure*}
    \centering
    \small
    \includegraphics[scale=0.95,trim={0cm 0cm 0cm 0cm},clip]{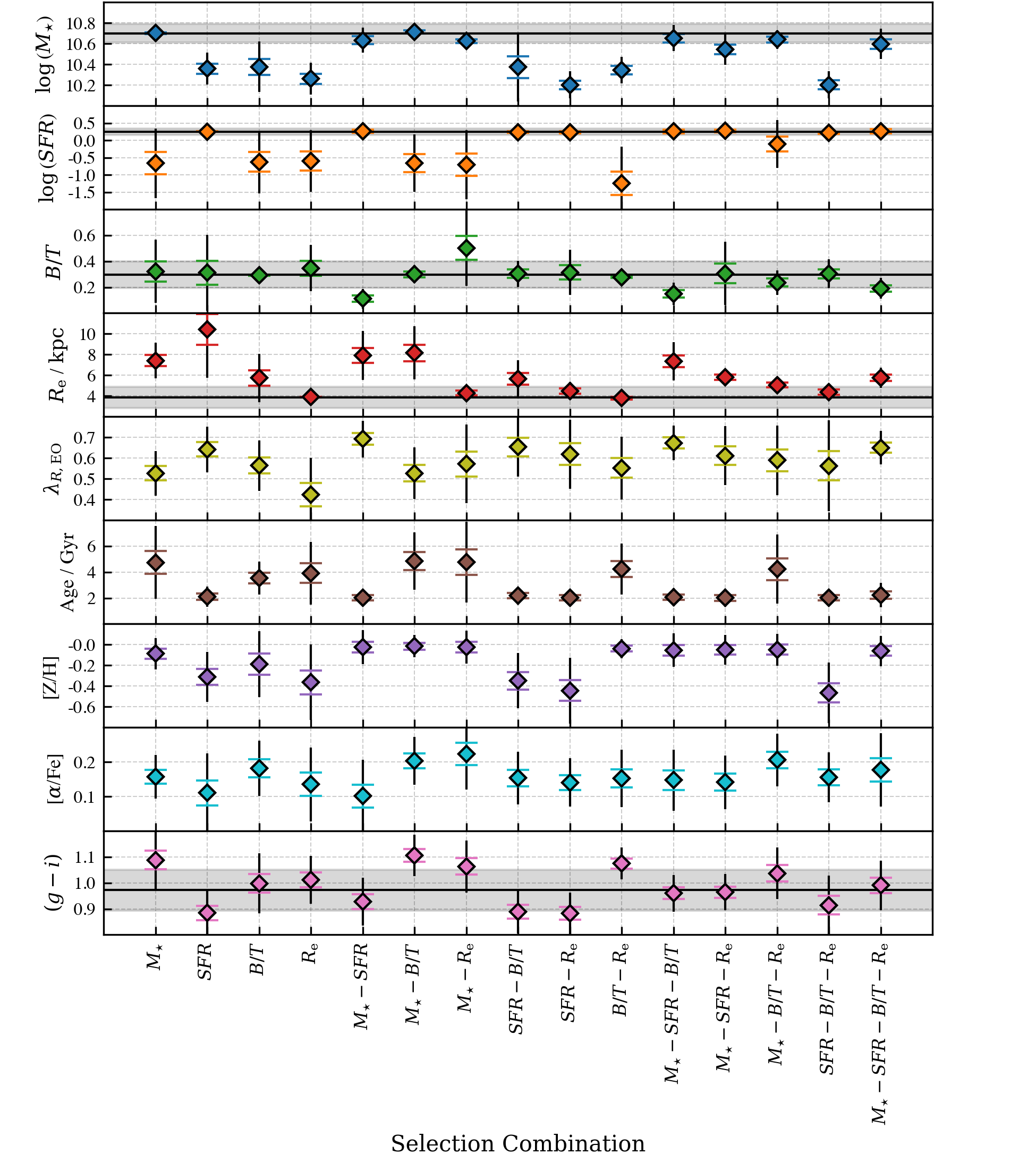}
    \caption{Comparisons between the averaged properties of the top-10 GAMA MWAs selected by 15 parameter combinations. There are nine properties from top to bottom including $\log(M_{\star})$ (blue), $\log(SFR)$ (orange), $B/T$ (green), $R_{\rm{e}}$ (red), $\lambda_{R,\rm{EO}}$ (olive), age (brown), metallicity $[\rm{Z}/\rm{H}]$ (purple), $\alpha-$abundance $[\alpha/\rm{Fe}]$ (cyan) and rest-frame colour $(g-i)$ (pink). All diamond markers are accompanied by standard deviations (black error bars) and error on the mean (coloured caps). The Milky Way parameter in each diagram is shown as a black solid line with $1\sigma$ grey shaded area except $\lambda_{R,\rm{EO}}$, age, $[\rm{Z}/\rm{H}]$ and $[\alpha/\rm{Fe}]$. These comparisons show that each parameter combination provides MWAs with different properties. We find that some averaged properties are inconsistent with those of the Milky Way when this specific parameter is excluded from the selection combination.}
    \label{fig:Averages_GAMA_plots}
\end{figure*}

\begin{figure*}
    \centering
    \small
    \includegraphics[scale=0.9,trim={1.3cm 1.0cm 5.0cm 2.5cm},clip]{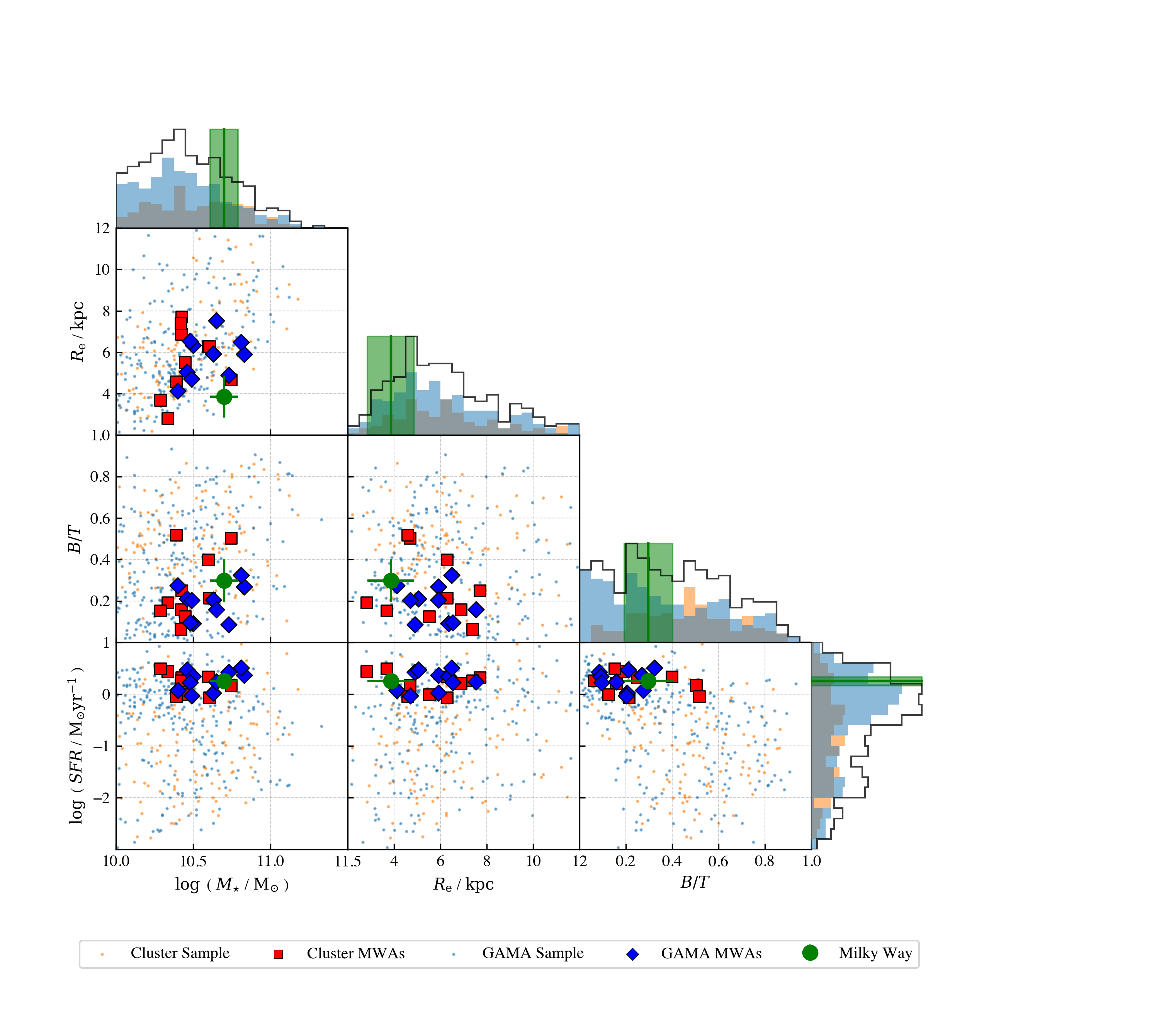}
    \caption{The distributions of SAMI galaxies after flagging (as described in Sec. \ref{subsec:flagging}) as a function of stellar mass $\log(M_{\star})$, star formation rate $\log(SFR)$, bulge-to-total ratio $B/T$ and disk effective radius $R_{\rm{e}}$. The green dots with error bars represent the Milky Way parameters within $1\sigma$. The light blue and orange dots represent the GAMA and Cluster samples, respectively. The blue diamonds and red squares mark the GAMA and Cluster top-10 MWAs, respectively. Histograms illustrate the marginal distributions for the SAMI galaxies, with the GAMA and Cluster SAMI galaxies in light blue and orange shades. The Milky Way parameters are marked on the histograms as green lines with $1\sigma$ green shades. The histograms demonstrate that the Milky Way's stellar mass and disk effective radius are significantly different from the majority of the SAMI galaxies.}
    \label{fig:corner_plot}
\end{figure*}

We calculate all averaged properties and accompanied standard deviations and illustrate the numerical results in Appendix \ref{appendix:MWAs_prop}. In Fig. \ref{fig:Averages_GAMA_plots} we present the averaged properties of the selected MWAs including $\log(M_{\star})$, $\log(SFR)$, $B/T$ and $R_{\rm{e}}$ from each selection combination for the GAMA region following the results in Table \ref{tab:GAMA_MWAs}.

When we include a selection parameter into a selection combination, this averaged property of the top-10 MWAs will agree with the Milky Way within $1\sigma$. On the other hand, an averaged property either disagrees within $1\sigma$ or has a large standard deviation when we exclude the selection parameter from a selection combination. For instance, all selections without $R_{\rm{e}}$ provide MWAs with large averaged $R_{\rm{e}}$ values. The selections excluding $\log(SFR)$ always push the MWAs to have lower averaged SFRs as there are few galaxies with higher SFRs than the Milky Way at a similar stellar mass. Unlike other parameters, $B/T$ is not significantly biased by the other selections. This agreement/disagreement encourages the usage of as many selection parameters as possible to ensure that all properties of the MWAs are close to those of the Milky Way.

Despite the agreement within $1\sigma$ with the Milky Way's properties, there are biases in terms of $\log(M_{\star})$ and $R_{\rm{e}}$. The biases towards lower stellar mass and greater disk effective radius are expected due to the shape of the stellar mass-size function in the samples, i.e. there are more low-mass galaxies than high-mass galaxies and the size of the Milky Way is more compact than other galaxies with its stellar mass. When we select MWAs with four selection parameters, we find that the top-10 GAMA MWAs have 0.10 dex lower average stellar mass and 1.9 kpc greater average disk effective radius than the Milky Way. The Cluster MWAs have 0.23 dex lower average stellar mass and 1.7 kpc greater average disk effective radius than the Milky Way.

To understand the effects of selecting only a small number of MWAs, we increase the number of MWAs from 10 to 50 and repeat our analysis. At this number, the MWAs account for approximately 20\% and 37\% in the GAMA and Cluster regions, respectively. Unsurprisingly, we find that the averaged properties of the top-50 MWAs moved towards the averaged properties of the entire samples. Hence, the selection with a large number of MWAs would rather represent the entire sample instead of the Milky Way-like galaxies. We explain the changes in the averaged properties when increasing the number of galaxies being selected as MWAs in Appendix \ref{appendix:bias}.

Finally, we decided to utilise the selection based on all four selection parameters. This is because all four averaged properties of the selected MWAs, including $\log(M_{\star})$, $\log(SFR)$, $B/T$ and $R_{\rm{e}}$, agree within $1\sigma$ as illustrated in Fig. \ref{fig:Averages_GAMA_plots}. All of the top-10 selected MWAs from the GAMA and Cluster regions are located near the Milky Way as shown in Fig. \ref{fig:corner_plot}.

\section{The properties of MWAs}
\label{sec:MWAs_prop}

We obtain the top-10 MWAs selected using all four selection parameters as illustrated in Fig. \ref{fig:corner_plot}. In this section we describe the properties of these MWAs including morphology, stellar kinematics, star formation rate and mean stellar population properties, rest-frame $(g-i)$ colour and environments. We compare the properties of the MWAs to those of the Milky Way where appropriate.

\begin{figure*}
    \centering
    \includegraphics[scale=0.94,trim={0cm 0.5cm 0cm 0.25cm},clip]{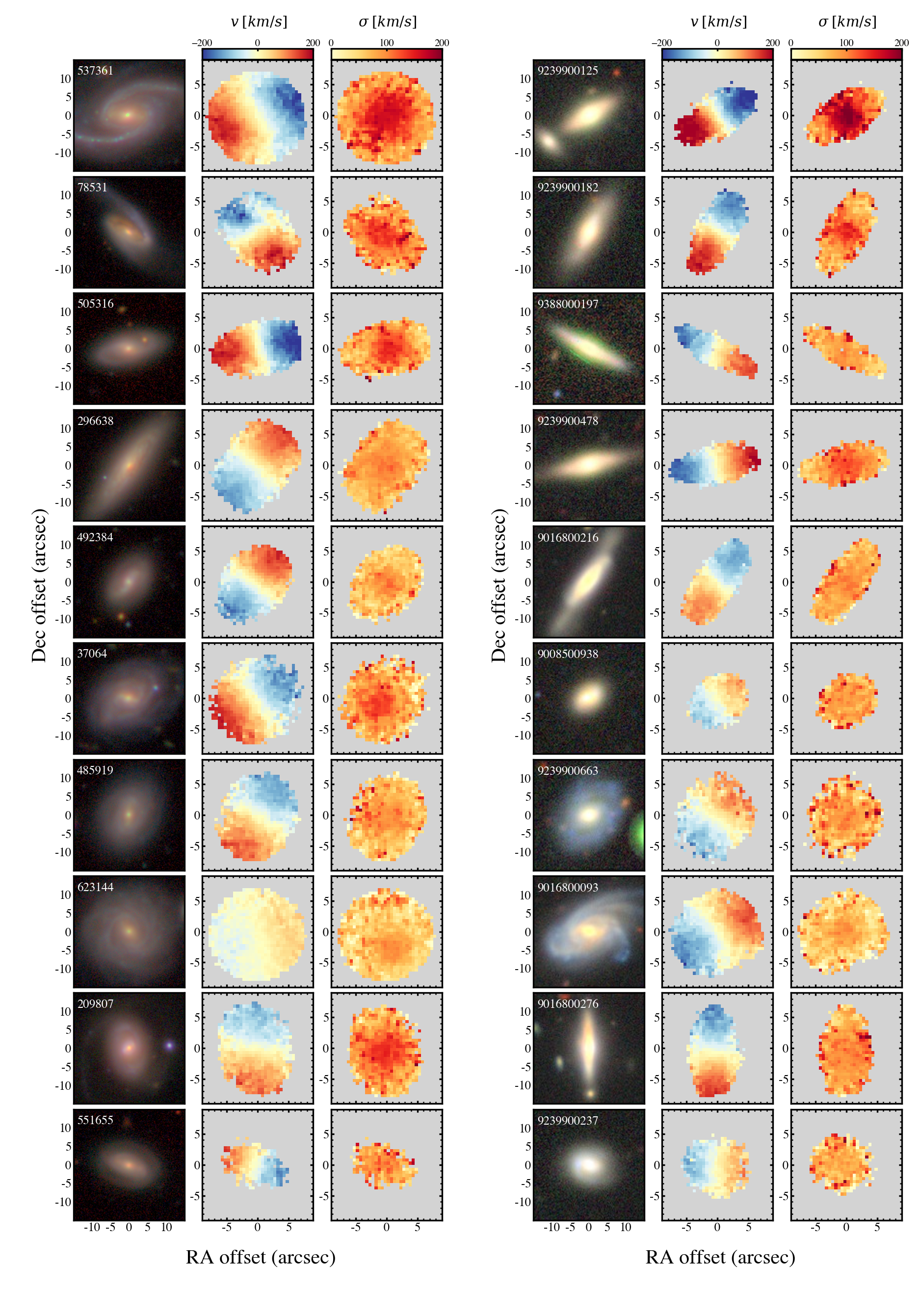}
    \caption{The images, stellar velocity maps and stellar velocity dispersion maps of top-10 MWAs selected using $M_{\star}$, $SFR$, $B/T$ and $R_{\rm{e}}$ criteria from GAMA and Cluster regions on the left and right columns, respectively. The images of GAMA MWAs come from the KiDS \citep{Kuijken2019}, while the images of Cluster MWAs come from the DECaLS \citep{Dey2019}. The colours of the stellar velocity maps indicate the stellar motion either moving towards (blue) or away (red) from the observers. The colour coding in the stellar velocity dispersion maps indicates the randomness of the stellar random motion from low to high. The images show that the MWAs are disk galaxies with a small bulge component and some of them show a bar component. The stellar velocity and velocity dispersion show that the MWAs are rotationally supported and usually have centrally-peaked velocity dispersion profiles.}
    \label{fig:mapping}
\end{figure*}

\subsection{Morphology}

Fig. \ref{fig:mapping} illustrates the morphology of the MWAs through RGB images. The images of the GAMA MWAs are from the KiDS \citep{Kuijken2019} with a pixel size of $0.20$ arcsec and seeing of $0.70$ arcsec in $r-$band. The images of the Cluster MWAs are from the DECaLS \citep{Dey2019} with a pixel size of $0.27$ arcsec and seeing of $1.30$ arcsec in $r-$band. Based on the visual morphology from the SAMI catalogue \code{VisualMorphologyDR3} \citep{Cortese2016}, all GAMA MWAs and nine Cluster MWAs are classified as a spiral, disk galaxy and there is one Cluster MWA (galaxy 9239900125) classified as a lenticular galaxy. 

The MWAs have a prominent disk and a small bulge. The averaged $B/T$ of the GAMA and Cluster MWAs are $0.19 \pm 0.08$ and $0.26 \pm 0.15$, respectively. These $B/T$ are consistent with the estimated $B/T$ of the Milky Way at $0.30\pm0.10$. 

We calculate the probability of selecting a barred galaxy in a sample and derive its $1\sigma$ confidence interval using the method from \citet{Cameron2011}. We find that there are 137 and 48 barred galaxies in the GAMA and Cluster samples, which corresponds to bar fractions of $54\pm3\%$ and $36\pm4\%$. There are nine GAMA MWAs (except galaxy 78531) and two Cluster MWAs (galaxies 9239900663 and 9016800093) with a bar component which result in fractions of $90^{+3}_{-17}\%$ and $20^{+17}_{-7}\%$. The chance of having a GAMA MWA with a bar component exceeds the fraction in the whole GAMA sample. This means our MWA selection with four parameters preferentially selects GAMA MWAs with a bar component. On the other hand, the chances of picking a barred galaxy in the Cluster MWAs and the Cluster sample are comparable. A lack of barred Cluster MWAs is probably caused by the different image resolutions used for the classification. The detailed classification was done based on three databases including HSC \citep{Miyazaki2018}, DECaLS \citep{Dey2019} or PanSTARRS \citep{Flewelling2020}. The best quality images were from the HSC catalogue and unfortunately, not all the Cluster galaxies have coverage in the HSC sample. Hence, it is possible that the bar component could not be distinguished from other components and then identified as galaxies without bars.

\subsection{Stellar kinematics}

\begin{figure*}
    \centering
    \includegraphics[scale=0.55]{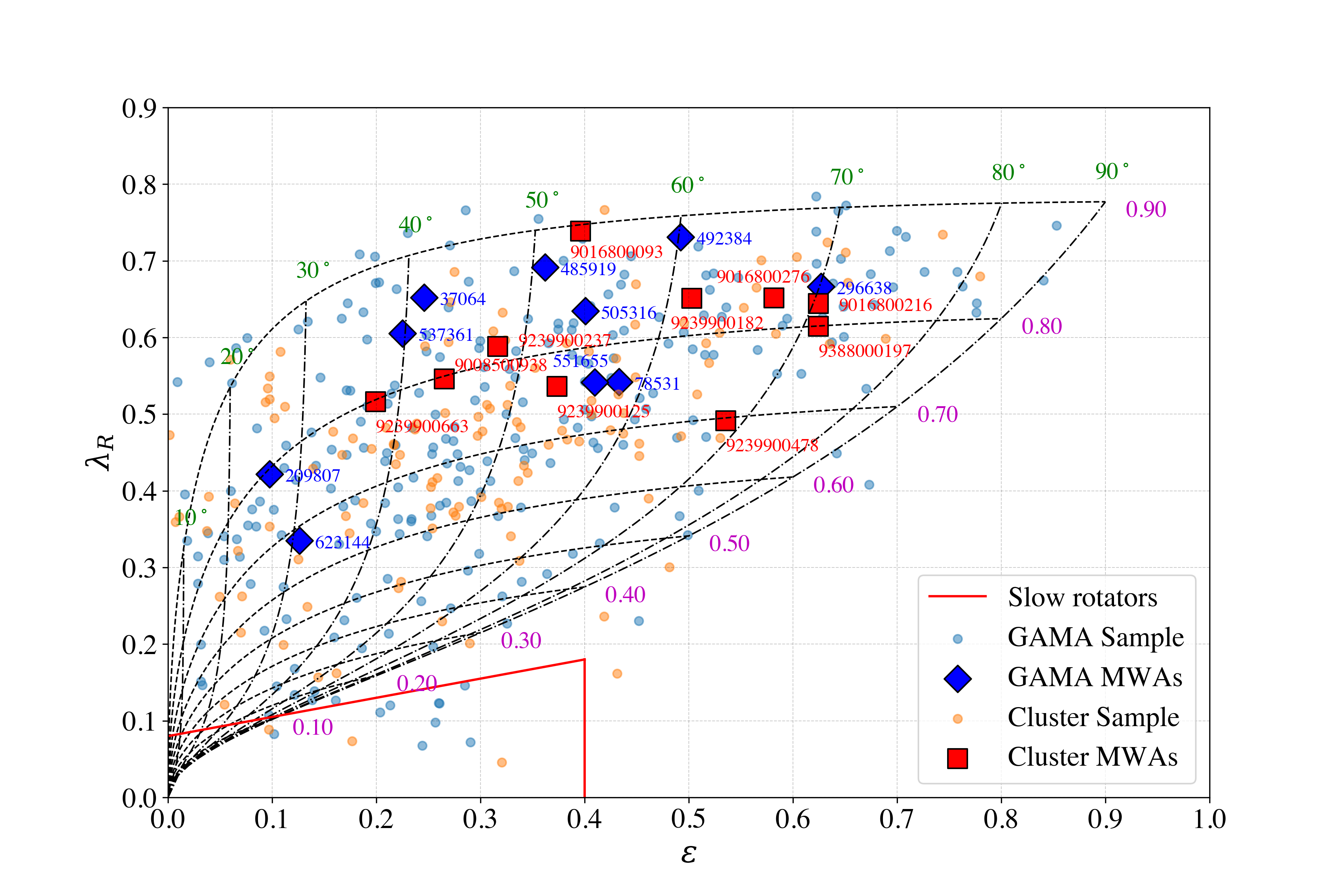}
    \caption{$\lambda_{R}$ versus $\varepsilon$, presenting the top-10 MWAs as large labelled markers among the samples with blue diamonds and red squares representing the GAMA and Cluster MWAs, respectively. The GAMA and Cluster samples are indicated as light blue and orange dots, respectively. The black curves illustrate the relationship between the $\lambda_{R}$ and $\varepsilon$ at the same inclination $i$ (vertically dot-dashed lines and green labels with $10^\circ$ intervals) and intrinsic ellipticity $\varepsilon_{i}$ (horizontally dashed lines and magenta labels with 0.1 intervals) \citep{Cappellari2007,Emsellem2011,Cappellari2016}. The red line shows the region of slow rotators as defined by \citet{Cappellari2007}. These results imply that all MWAs are fast-rotating galaxies with intrinsically thin disks.}
    \label{fig:ellip_lambdar}
\end{figure*}

The catalogue \code{samiDR3Stelkin} provides the stellar kinematics properties including velocity $v$, velocity dispersion $\sigma$, spin parameter $\lambda_{R}$, edge-on spin parameter $\lambda_{R,\rm{EO}}$, ellipticity $\varepsilon$, intrinsic ellipticity $\varepsilon_{i}$ and inclination $i$. We map the velocity and the velocity dispersion of the MWAs in Fig. \ref{fig:mapping} and locate the MWAs on the $\lambda_{R}-\varepsilon$ plane in Fig. \ref{fig:ellip_lambdar}.

The MWAs show a strong rotational component and an increasing dispersion towards the centre as seen in Fig. \ref{fig:mapping}. The results suggest that the majority of stars in the galaxies are in circular orbits and the velocity dispersion increases in the centres of the galaxies \citep{vandeSande2017b}. Fig. \ref{fig:ellip_lambdar} confirms that the MWAs are fast rotators and intrinsically thin disks. Unsurprisingly, none of them is located in the slow-rotator region of the $\lambda_{R}-\varepsilon$ diagram given the correlation between SFR and $\lambda_{R}$ that is \citep[e.g.][]{Guo2019,Cortese2022}.

We carry out statistical comparisons using the edge-on spin parameter $\lambda_{R,\rm{EO}}$ of the MWAs instead of the spin parameter $\lambda_{R}$ which depends on the inclination. The averaged edge-on spin parameters are $0.65 \pm 0.08$ and $0.64 \pm 0.06$ for the GAMA and Cluster MWAs, respectively. We show the averaged edge-on spin parameter of the GAMA MWAs from different selection combinations in Fig. \ref{fig:Averages_GAMA_plots}. The numerical outcomes are in Tables \ref{tab:GAMA_MWAs} in Appendix \ref{appendix:MWAs_prop}. There is a significant reduction in the averaged edge-on spin parameter when selecting MWAs only with $R_{\rm{e}}$ (i.e. $0.42\pm0.18$). The averaged edge-on spin parameters are marginally lower when excluding SFR from the selection combinations (i.e. approximately $0.5-0.6$). Fig. \ref{fig:ellip_lambdar} shows that the ellipticities $\varepsilon$ of the MWAs are varied between $0.1-0.7$. However, the intrinsic ellipticities $\varepsilon_{i}$ are typically high with means of $0.80 \pm 0.06$ and $0.81 \pm 0.05$ for the GAMA and Cluster MWAs, respectively. There is no significant difference in terms of stellar kinematics between the MWAs from GAMA and Cluster regions when the SFRs of galaxies are accounted for in the MWA selection \citep{Croom2024}.  

\subsection{Star formation and mean stellar population properties}

\begin{figure*}
    \centering
    \includegraphics[scale=0.38,trim={0.2cm 0cm 2.0cm 0cm},clip]{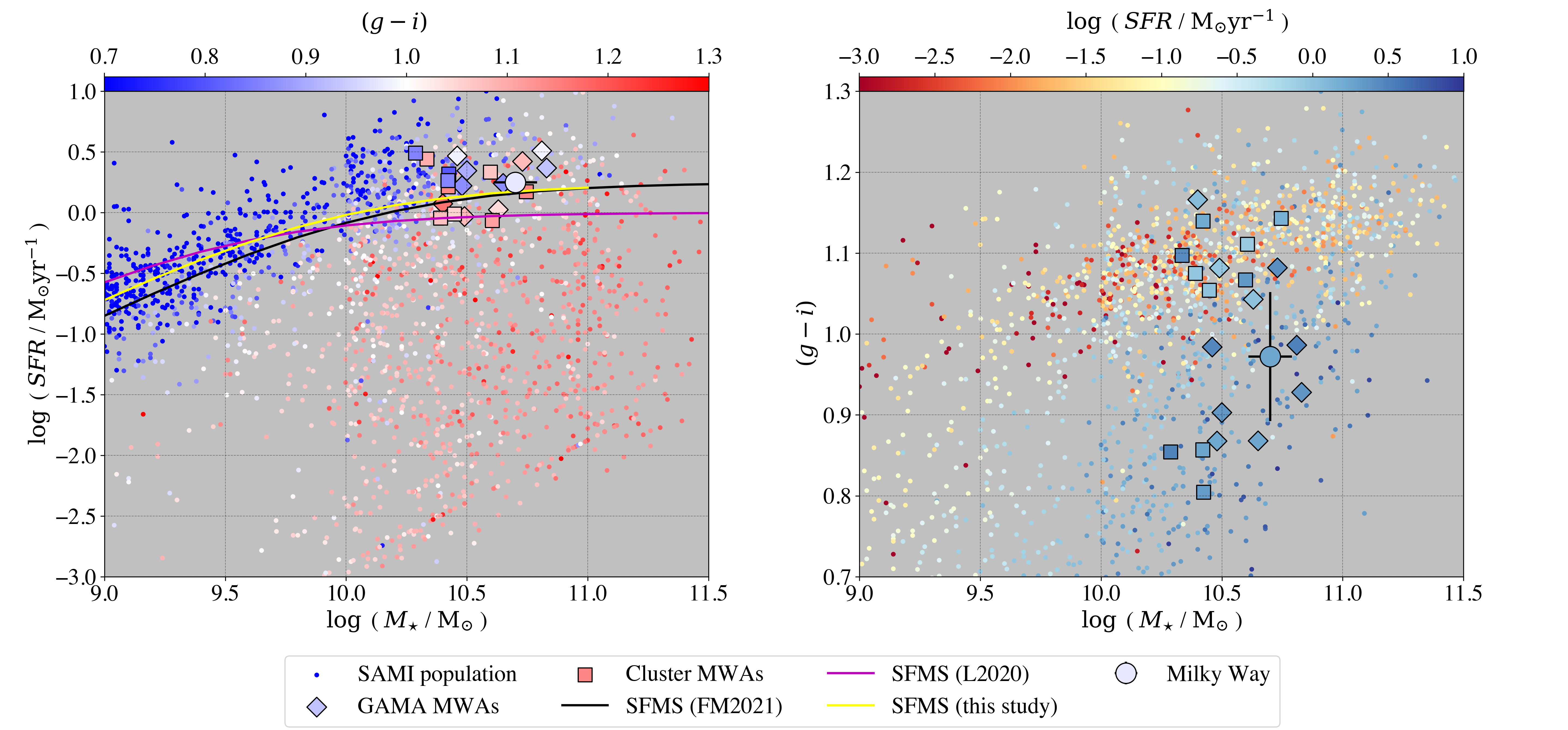}
    \caption{The correlation between the stellar mass $\log(M_{\star})$, the star formation rate $\log(SFR)$ and the rest-frame $(g-i)$ colour. All SAMI galaxies are shown as small dots. The MWAs are shown as diamonds for the GAMA MWAs and squares for the Cluster MWAs. The Milky Way is a large dot with $1\sigma$ error bars. \textit{(a) Left panel:} the star formation rates $\log(SFR)$ as a function of stellar mass $\log(M_{\star})$ colour coded by the rest-frame $(g-i)$ colour. The yellow solid curve illustrates the star-forming main sequence line (SFMS) calculated using Eq. \ref{eq:SFMS}. The black solid curve illustrates the SFMS calculated by \citet{FraserMcKelvie2021} which is consistent with the calculation in this paper. The magenta solid curve illustrates the SFMS calculated by \citet{Leslie2020}, based on a different set of SFR measurements. \textit{(b) Right panel:} the rest-frame $(g-i)$ colour as a function of stellar mass $\log(M_{\star})$ colour coded by the star formation rates $\log(SFR)$. This figure illustrates that the rest-frame $(g-i)$ colour of the MWAs are distributed within $3\sigma$ from the Milky Way at $0.972 \pm 0.080$ \citep{LNB2015,LN2016}.}
    \label{fig:M_SFR_gi_selbySFR}
\end{figure*}

\begin{figure*}
    \centering
    \includegraphics[scale=0.51,trim={0.25cm 0cm 1.5cm 0cm},clip]{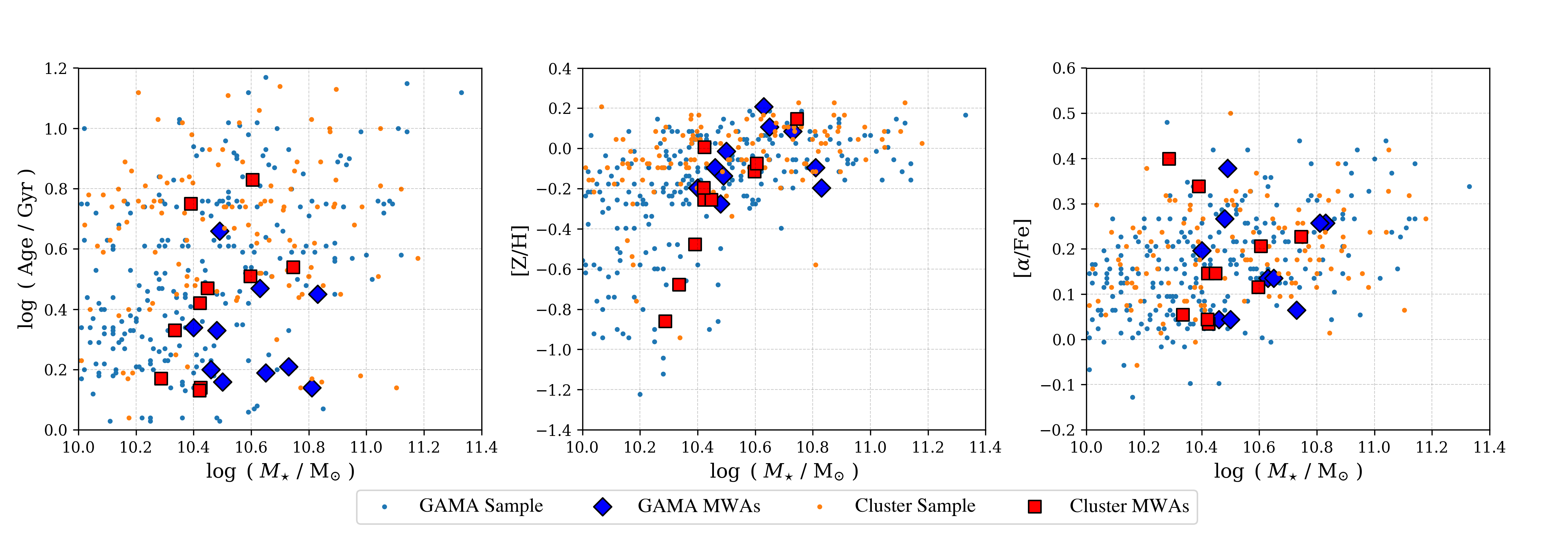}
    \caption{The light-weighted $\log(\rm{Age/Gyr})$ (left), metallicity $[\rm{Z}/\rm{H}]$ (middle) and $\alpha-$abundance $[\alpha/\rm{Fe}]$ as a function of stellar mass $\log(M_{\star})$ within one effective radius. The top-10 MWAs are the large blue diamonds and red squares representing the GAMA and Cluster MWAs, respectively. The GAMA and Cluster samples are indicated by small light blue and orange dots, respectively. The MWAs are younger than the mean of the population, as expected for a star-forming sample. Most MWAs have $[\rm{Z}/\rm{H}]$ and $[\alpha/\rm{Fe}]$ that is typical of the underlying population, but with a few Cluster MWAs scattering to lower $[\rm{Z}/\rm{H}]$.}
    \label{fig:M_Age_metallicity_alpha}
\end{figure*}

We define a star-forming main sequence (SFMS) from the SAMI galaxies in the GAMA regions following the method in \citet{FraserMcKelvie2021}. Within $9<\log(M_{\star}/\rm{M}_{\odot})<11$, we divide the galaxies in mass (each bin 0.2 dex) and determine the $\log(SFR)$ distribution for each of these bins. We consider the $\log(SFR)$ at the peak of the histogram as the location of the main sequence at that mass. We fit a relationship with a turnover at the high stellar mass of the form
\begin{equation}
    \log(SFR/\rm{M}_{\odot}\rm{yr}^{-1})=S_{0}-a_{1}t-\log\left(1+\frac{10^{M_{t}'}}{10^{M}}\right),
	\label{eq:SFMS}
\end{equation}
where $M$ is the midpoint of each $\log(M_{\star})$ bin \citep{Lee2015} and $t$ is the age of the universe in Gyr \citep[$t=13.5$ Gyr,][]{FraserMcKelvie2021}. The best fit values are $S_{0}=2.923$, $a_{1}=0.199$ and $M_{t}=9.909$. 

Figure \ref{fig:M_SFR_gi_selbySFR}(a) illustrates the $\log(SFR)$ of the selected MWAs as a function of $\log(M_{\star})$ together with the SFMS lines from Eq. \ref{eq:SFMS}, \citet{FraserMcKelvie2021} and \citet{Leslie2020} colour coded by the rest-frame $(g-i)$ colour. Note that our calculation refers to a smaller mass range due to a lack of high-mass star-forming galaxies in the SAMI catalogue. \citet{FraserMcKelvie2021} and \citet{Leslie2020} considered different galaxy samples which can lead to differences in their SFMS estimates. \citet{FraserMcKelvie2021} obtained the sample by combining data from SAMI and MaNGA Surveys with a mass range of $9<\log(M_{\star}/\rm{M}_{\odot})<11.7$. While \citet{Leslie2020} determined the SFMS from radio-continuum galaxies in the VLA-COSMOS 3 GHz Large Project \citep{Smolcic2017} with a mass range of $8.5<\log(M_{\star}/\rm{M}_{\odot})<11.5$. As expected, the SFMS is linear at low stellar mass ($\log(M_{\star}/\rm{M}_{\odot})<10$) but flattens at high stellar mass \citep[e.g.][]{Leslie2020,FraserMcKelvie2021}. The Milky Way is located slightly above all three SFMS trends, while the MWAs from both GAMA and Cluster regions are located within $\pm0.4$ dex of the derived SFMS.

Figure \ref{fig:M_Age_metallicity_alpha} illustrates light-weighted age $\log(\rm{Age/Gyr})$, metallicity $[\rm{Z}/\rm{H}]$ and $\alpha-$abundance $[\alpha/\rm{Fe}]$ as a function of of $\log(M_{\star})$. We show the averaged age, $[\rm{Z}/\rm{H}]$ and $[\alpha/\rm{Fe}]$ of the GAMA MWAs from different selection combinations in Fig. \ref{fig:Averages_GAMA_plots}. The numerical outcomes are shown in Tables \ref{tab:GAMA_MWAs} in Appendix \ref{appendix:MWAs_prop}. The averaged ages and their standard deviations of the top-10 GAMA and Cluster MWAs are $2.2\pm0.9$ Gyr and $3.1\pm1.7$ Gyr, respectively, which imply young light-weighted stellar population. The estimated ages of the MWAs would be higher if SFR is excluded from the selection. The averaged metallicities of the top-10 GAMA and Cluster MWAs are $-0.06\pm0.15$ and $-0.28\pm0.30$, respectively. The Cluster MWAs have a wider range of metallicities than the GAMA MWAs. The estimated metallicities would be lower if stellar mass is excluded from the selection. The averaged $\alpha-$abundances of the top-10 GAMA and Cluster MWAs are $0.18\pm0.11$ and $0.17\pm0.12$, respectively. The $\alpha-$abundance is not significantly biased by the other selections. The MWAs are located within the same ranges of metallicities and $\alpha-$abundances as the parent samples.

\subsection{The rest-frame (g-i) colour}

Fig. \ref{fig:M_SFR_gi_selbySFR}(b) illustrates the rest-frame $(g-i)$ colour of the SAMI population as a function of stellar mass $\log(M_{\star})$. The estimated rest-frame $(g-i)$ colour of the Milky Way is $0.972 \pm 0.080$ calculated from MWAs based on SFRs \citep{LNB2015,LN2016}. The averaged rest-frame $(g-i)$ colours of the GAMA and Cluster MWAs are $0.99\pm0.10$ and $1.02\pm0.12$, respectively, which are consistent with the \citet{LN2016} estimate of the Milky Way colour. The Milky Way can be categorised as a green valley member as it resides in-between the blue and red colour populations \citep{Mutch2011}. The MWAs are scattered within $3\sigma$ from the Milky Way's $(g-i)$ value and are found either in the blue cloud, green valley, or red sequence.

We show the averaged rest-frame $(g-i)$ colour of the GAMA MWAs from different selection combinations in Fig. \ref{fig:Averages_GAMA_plots}. The numerical outcomes are in Tables \ref{tab:GAMA_MWAs} in Appendix \ref{appendix:MWAs_prop}. All selections obtain averaged rest-frame $(g-i)$ colours of the GAMA MWAs that agree with the estimate from \citet{LNB2015,LN2016}. For the selections of the Cluster MWAs, however, the averaged rest-frame $(g-i)$ colour of the galaxies selected by $M_{\star}-R_{\rm{e}}$ and $M_{\star}-B/T-R_{\rm{e}}$ disagree with the Milky Way value. Given the large number of passive galaxies in the clusters, ignoring the SFR as a selection parameter leads to greater biases.

As the colours of galaxies correlate with their SFRs, it might be plausible to substitute the rest-frame $(g-i)$ colour for the SFR in the MWA selection. However, if we replace the SFR with the rest-frame $(g-i)$ colour, the averaged SFR of the GAMA MWAs is $-0.04\pm0.66$ $\rm{M}_{\odot}\rm{yr}^{-1}$, which is significantly lower than that of the Milky Way. This result implies that we should be careful to substitute the SFR with the rest-frame $(g-i)$ colour for selecting MWAs because the correlation between the rest-frame $(g-i)$ colour and the SFR has a large scatter. We describe the derivation of the rest-frame $(g-i)$ colours of the Milky Way and the SAMI galaxies together with the selection of MWAs based on the selection combination $M_{\star}-(g-i)-B/T-R_{\rm{e}}$ in Appendix \ref{appendix:(g-i) colour}.

\subsection{Environments}

\begin{figure*}
    \centering
    \includegraphics[scale=0.65,trim={0.0cm 0.0cm 0.2cm 0cm},clip]{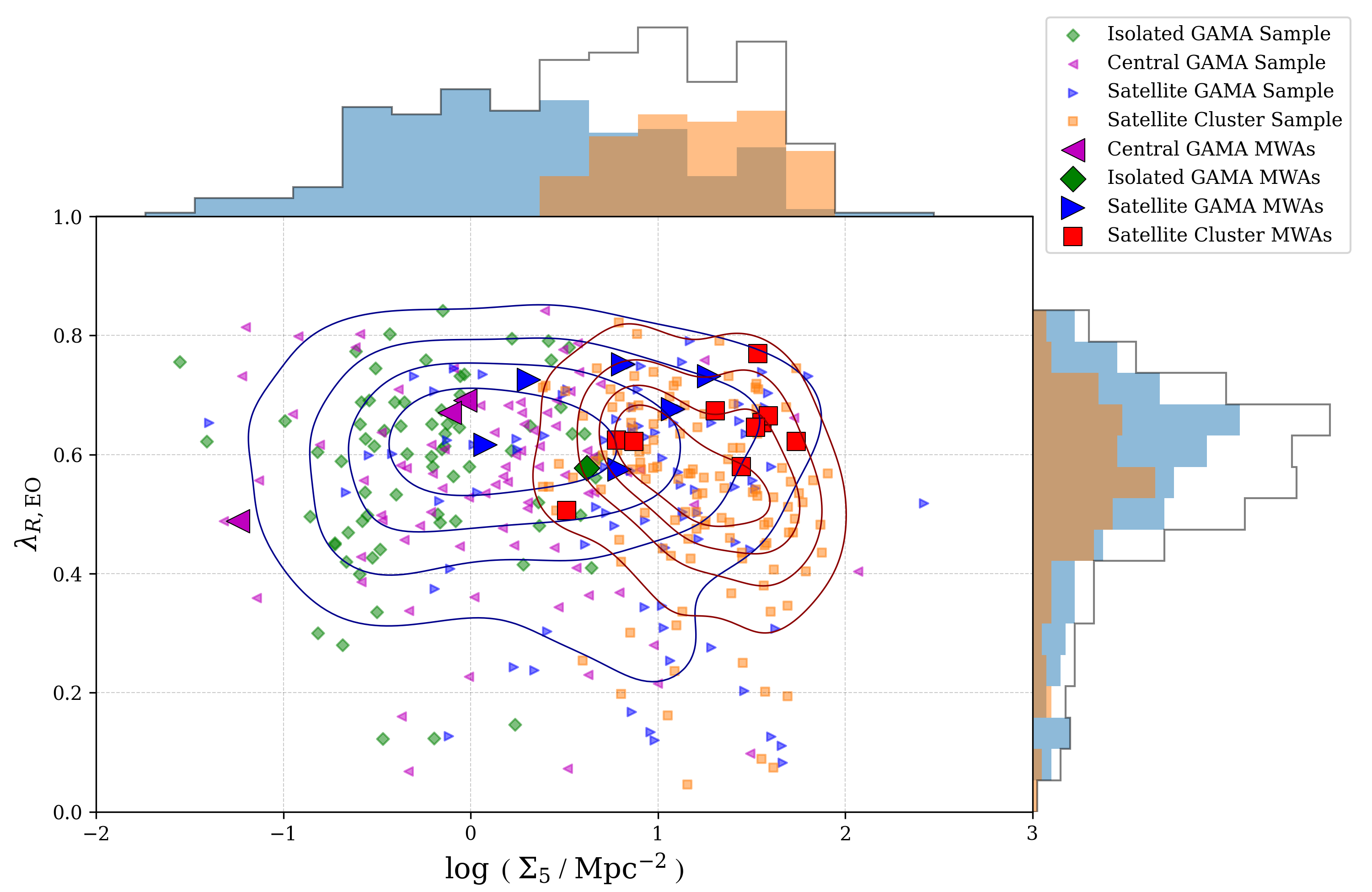}
    \caption{The edge-on spin parameter $\lambda_{R,\rm{EO}}$ as a function of surface density $\Sigma_{5}$. We show the GAMA and Cluster samples as small markers: green diamonds, magenta left triangles and blue right triangles for isolated, central and satellite GAMA galaxies, respectively, and orange squares for Cluster galaxies. We separate the GAMA MWAs into isolated (large green diamond), central (large magenta left triangles) and satellite (large blue right triangles) galaxies. All of the Cluster MWAs are satellite galaxies (large red squares). Histograms illustrate the marginal distributions for the GAMA and Cluster samples in light blue and orange shades, respectively. The histograms for the total samples are drawn with grey lines. The contours show the kernel-density estimates for the separate distributions of the GAMA (dark blue) and Cluster (dark red) samples, respectively. This figure illustrates that the MWAs have comparably high edge-on spin parameters with higher average surface densities in the Cluster regions.}
    \label{fig:Sigma5_Spin}
\end{figure*}

We apply the MWA selection separately to the galaxies in the GAMA and Cluster regions to investigate the influences of environments. We find that the stellar mass, SFR, $B/T$, disk effective radius, morphology, stellar kinematics, mean stellar population properties and $(g-i)$ colour of the GAMA and Cluster MWAs are comparable despite the different environments. In other words, we can find MWAs in Cluster environments when using an identical selection technique to that used in the GAMA regions. Even though the distributions of the Cluster MWAs in Fig. \ref{fig:corner_plot} are slightly broader than those of the GAMA MWAs, the averaged properties of the Cluster MWAs agree with those of the Milky Way within $1\sigma$. We present the averaged properties of the Cluster MWAs in Table \ref{tab:cluster_MWAs}.

We determine an environmental density using the surface density. The surface density was estimated using the projected comoving distance to the fifth nearest neighbour, which was drawn from a density-defining population of galaxies with absolute magnitudes $M_{r} < -18.6$ magnitude and within a velocity $\pm1000$ km s$^{-1}$ with respect to the redshift of the SAMI target of interest \citep{Brough2013,Brough2017}. The surface density of the SAMI population was listed in the catalogue \code{DensityCatDR3} \citep{Croom2021}. We also consider whether the MWAs are in galaxy groups using the data listed in the catalogue \code{G3CFoFGroup} \citep{Robotham2011} and divide galaxies according to whether they are an isolated galaxy (i.e. the galaxy is not in a galaxy group), a central galaxy, or a satellite galaxy.

Fig. \ref{fig:Sigma5_Spin} shows the edge-on spin parameter $\lambda_{R,\rm{EO}}$ as a function of surface density $\Sigma_{5}$. This agrees with \citet{Croom2024} who shows that there is little independence of spin parameter on environments once stellar population age or SFR is accounted for. The MWAs have high edge-on spin parameters independent of the environmental density. All of the Cluster MWAs and five GAMA MWAs are satellite galaxies located in regions with surface densities between $0<\log(\Sigma_{5}/\rm{Mpc}^{-2})<2$. There are three GAMA MWAs (galaxies 537361, 505316 and 623144) that were categorised as central galaxies of galaxy groups. These central galaxies are situated in less dense environments with $-2<\log(\Sigma_{5}/\rm{Mpc}^{-2})<0$. There is only one GAMA MWA (galaxy 551655) that was categorised as an isolated galaxy with $\log(\Sigma_{5}/\rm{Mpc}^{-2})\sim0.62$. This high surface density was caused by a reasonable number of galaxies sitting close in projection, but behind 551655, just within the $\pm1000$\,km\,s$^{-1}$ window used for the $\Sigma_5$ calculation. The group matching in \citet{Robotham2011} was done in three dimensions and the isolated GAMA MWA is sufficiently offset in velocity to not be considered as a part of a galaxy group.

We find only two GAMA MWAs with evidence of interactions: galaxy 537361 is a central, barred galaxy and galaxy 78531 is a satellite, non-barred galaxy. There are two satellite GAMA MWAs (37064 and 209807) without ongoing interactions which have spiral central galaxies in their galaxy groups with a slightly larger stellar mass and a smaller likelihood of having a bar component than their MWAs (the parameter \code{tot\_bar\_fract} of the spiral central galaxy is less than that of its satellite MWA). These systems could arguably be Milky Way-M31 analogues. We provide the images of the galaxy groups with GAMA MWAs in Appendix \ref{appendix: companions}.

There are three Cluster MWAs plausibly having ongoing interactions: galaxies 9016800216, 9016800093 and 9239900237. These Cluster MWAs have shells or asymmetry which are likely merger remnants. We consider whether there is a link between the existence of interactions and location within a cluster using the parameter $R/R_{200}$ which is the ratio between the projected distance from the cluster centre and the virial radius of the cluster \citep{Owers2017}. If a galaxy is located within $R/R_{200}<1$, environmental effects can significantly impact the satellite galaxies \citep{Wetzel2012}. We find that all three Cluster MWAs with possible ongoing interactions have an averaged $R/R_{200}$ of $0.46\pm0.08$. While the other seven Cluster MWAs without evidence for interactions have an averaged $R/R_{200}$ of $0.74\pm0.33$. The difference in the averaged $R/R_{200}$ suggests that the Cluster MWAs with possible ongoing interactions locate closer to the centres of the clusters than those without evidence of interactions. However, this difference is not statistically significant.

\section{Discussion}
\label{sec:discussion}

This section discusses the implications of using different selection methods for finding and analysing MWAs, placing our work in context with previous studies.

\subsection{Selection Strategy}

Three main selection methods have been previously used to define MWAs. The first is based on selecting MWAs within $1\sigma$ of key Milky Way parameters. \citet{FraserMcKelvie2019} selected galaxies that fall within $10.69<\log(M_{\star}/\rm{M}_{\odot})<10.86$ and $0.131<B/T<0.178$ and found only $\sim 0.01\%$ of galaxies meeting these criteria. Similarly, \citet{Boardman2020a} used $10.66<\log(M_{\star}/\rm{M}_{\odot})<10.86$, $0.13<B/T<0.19$ and $1.46<SFR/\rm{M}_{\odot}\rm{yr}^{-1}<1.84$ and found only $\sim 0.001\%$ of galaxies. While this selection method ensures that these MWAs are as close to the true Milky Way values as possible, this method becomes impractical for finding a statistically representative sample of MWAs in surveys such as SAMI and MaNGA as only a small number of galaxies meets these strict selection criteria.

Another method pairs two selection parameters and then identifies MWAs from a joint probability distribution function \citep{LNB2015,Boardman2020a}. This selection method results in MWAs with less strict properties, hence the number of the MWAs would be larger than that from the first method. However, in our work, we find that using only two parameters can include significant biases in other parameters. For example, when selecting galaxies on stellar mass and bulge-to-total ratio, we find that the averaged disk effective radius of the selected MWAs is too high and the averaged SFR is too low (Fig. \ref{fig:Averages_GAMA_plots}).

The third method is based on the nearest neighbours approach which is applied in \citet{Pilyugin2023} and this study. This method calculates a distance $D$, from the Milky Way to galaxies in the multi-dimensional parameter spaces of selection parameters and sorts the distances of the galaxies. The nearest neighbours method allows us to customise the number of selection parameters as well as the number of galaxies being identified as MWAs. The number of Milky Way-like objects depends on the sample size. In particular, the number of MWAs depends on the defined distance threshold, which, when increased, results in the averaged properties of the MWAs getting closer to the averaged properties of the parent sample.

One challenge with all selection methods is the bias that is created by having a galaxy population that is unevenly distributed throughout the parameter space. In this work, the biases are illustrated in Fig. \ref{fig:corner_plot} where our selected MWAs are weighted towards lower stellar masses and larger disk effective radii, as there are more galaxies in these regions of parameter spaces. 

The MWA selections can be based on various parameters. The first selection parameter is usually stellar mass $M_{\star}$ as most properties of a galaxy strongly correlate with it. There are several studies applying the stellar mass as the only selection parameter such as MWA studies in simulations \citep[e.g.][]{vanDokkum2013,Karunakaran2021,SegoviaOtero2022,Font2022,Gurvich2022,MiroCarretero2023,Pan2023}. There may be a requirement of additional selection parameter(s) including SFR, colour, $B/T$, morphology and size when studying MWAs for specific analyses such as stellar kinematics, stellar population, ionised gas contents, environments, chemical evolution and star formation history. A large number of studies have employed the two-parameter selection technique such as $M_{\star}-SFR$ \citep{LNB2015,vandeSande2024}, $M_{\star}-B/T$ \citep{Boardman2020a}, $M_{\star}-$morphology \citep{AragonCalvo2023,Renaud2022,OrtegaMartinez2022,Lu2022} and $M_{\star}-$size \citep{Pilyugin2023}. Other studies used three parameters including $M_{\star}-SFR-$morphology \citep{Guszejnov2017}, $M_{\star}-B/T-$morphology \citep{FraserMcKelvie2019,Zhou2023} and $M_{\star}-$colour$-$morphology \citep{Krishnarao2020b,Scott2021}. 

In our study, we increase the number of selection parameters to four by adding the size parameters to the selection combination. The distribution of our selection on the $\log(M_{\star})-\log(SFR)$ plane in Fig. \ref{fig:M_SFR_all} illustrates that not including SFR as a selection parameter will pick out some passive galaxies. In terms of size, Fig. \ref{fig:M_SFR_all}(o) shows that the MWAs selected with size ($\log(M_{\star})-\log(SFR)-B/T-R_{\rm{e}}$) are similar to the selections without size ($\log(M_{\star})-\log(SFR)-B/T$) as seen in Fig. \ref{fig:M_SFR_all}(k). Therefore, the size parameter only marginally affects the choices of stellar masses and SFRs of the MWAs (e.g. Table \ref{tab:GAMA_MWAs} and \ref{tab:cluster_MWAs}). On the other hand, Fig. \ref{fig:BT_RE_all}(o) shows that our selection restricts the disk effective radius of the MWAs to be within $2.8-7.7$ kpc, whereas the size of the MWAs selected without size in Fig. \ref{fig:BT_RE_all}(k) ranges between $4.7-13.0$ kpc which biases to the larger disk effective radii. In terms of $B/T$, Fig. \ref{fig:BT_RE_all} shows that this is the least helpful selection parameter, as all selections provide MWAs within $1\sigma$ of the Milky Way $B/T$.

In summary, we apply four selection parameters simultaneously, stellar mass $M_{\star}$, star formation rate $SFR$, bulge-to-total ratio $B/T$ and disk effective radius $R_{\rm{e}}$ to maximise the level of the Milky Way-likeness for several aspects and provide MWAs that are appropriate for a wide range of topics. We confirm that combining the size with other selection parameters while selecting MWAs can provide the most restrictive candidates for MWAs \citep{Boardman2020a,Boardman2020b,Pilyugin2023}.

\subsection{The properties of the MWAs}

The averaged stellar masses, star formation rates, bulge-to-total ratios and disk effective radii of the top-10 MWAs from the GAMA and Cluster regions are within $1\sigma$ of the values of the Milky Way. Nonetheless, the MWAs selected in the GAMA regions have a smaller average distance $D$ from the Milky Way as compared to MWA selected from the Cluster regions. Fig. \ref{fig:corner_plot} shows that the top-10 selected MWAs in the GAMA and Cluster regions are biased towards lower stellar masses and have larger disk effective radii than the Milky Way. Our results confirm the previous work of \citet{LN2016} who showed that the disk of the Milky Way is remarkably small for its stellar mass, with a disk effective radius of $3.9\pm1.0$ kpc compared to $7.4\pm1.7$ kpc galaxies of similar mass. Nonetheless, when using our four selection parameters, the average disk effective radii of the top-10 MWAs are within $1\sigma$ from the values of the Milky Way (Fig. \ref{fig:Averages_GAMA_plots} and Fig. \ref{fig:Averages_Cluster_plots}). In contrast, the $B/T$ is not significantly biased in any by the selection combinations because the $B/T$ of the MWAs is within $1\sigma$ of the mean $B/T$ of the SAMI samples. 

The visual morphology and the stellar kinematics of our selected MWAs are correlated with the SFR. Applying the SFR to the MWA selection results in star-forming MWAs which are likely to have a rotationally-dominated disk with a bar component and a young stellar population \citep{vandeSande2018}. Fig. \ref{fig:mapping} shows that most of the GAMA MWAs have a clear disk with a bar component. Fig. \ref{fig:ellip_lambdar} illustrates that all of the MWAs are disk-dominated and rotationally-supported. Similar to \citet{Boardman2020a}, we find that our MWAs are consistent with being thin disks with an averaged intrinsic ellipticity of approximately 0.8. We can estimate the edge-on spin parameter $\lambda_{R,\rm{EO}}$ of the Milky Way based on the selected MWAs lying between $0.6-0.7$, this estimation will decrease if we exclude SFR from the MWA selections (Fig. \ref{fig:Averages_GAMA_plots}).

Fig. \ref{fig:M_Age_metallicity_alpha} indicates that as expected the MWAs are star-forming galaxies with young light-weighted stellar populations. We can estimate the age within 1 $R_{\rm{e}}$ of the Milky Way based on the selected GAMA MWAs lying between $1.4-4.6$ Gyr which is consistent with the estimation in \citet{Boardman2020a}. The metallicities and $\alpha-$abundances of the MWAs are typical of the underlying populations. We can estimate the metallicity and $\alpha-$abundance within 1 $R_{\rm e}$ of the Milky Way based on the selected GAMA MWAs lying between $-0.4\lesssim[\rm{Z}/\rm{H}]\lesssim0.2$ and $0.0\lesssim[\alpha/\rm{Fe}]\lesssim0.4$, respectively.

We find that the MWAs are located across a wide range of environmental surface densities (i.e. $\Sigma_{5}$) without significant difference between MWA selected from the GAMA and Cluster regions, consistent with the work from \citet{Croom2024}. Fig. \ref{fig:Sigma5_Spin} shows that the satellite MWAs are located within denser environments than central and isolated MWAs. Some MWAs have ongoing interactions which may be caused by the relatively small distance between galaxies in groups. The GAMA MWAs that are satellites and have a spiral galaxy as the central group member (i.e. the Milky Way-M31 analogues) show no signs of ongoing interactions. For the Cluster MWAs, we note that 3 out of 10 galaxies are located within the cluster virial radii and have ongoing interactions. We deduce that applying our MWA selection method provides an opportunity to explore the MWAs in various environments possibly as well as finding MWAs with ongoing interactions.

Overall, based on our MWA selection in the GAMA and Cluster regions, the averaged properties of the MWAs including stellar mass, SFR, $B/T$ and disk effective radius agree on the values of the Milky Way within $1\sigma$. The visual morphology of the majority of the MWAs is consistent with the Milky Way as a barred, spiral galaxy. We can predict the mean stellar kinematic properties of the Milky Way to be consistent with an intrinsically thin disk galaxy and a rotationally-dominated disk with a central high-velocity dispersion region. We can also predict the mean stellar population properties of the Milky Way to be consistent with a star-forming galaxy with young light-weighted stellar population. The differences in the properties of the GAMA and Cluster MWAs including morphology, stellar kinematics, stellar population properties and rest-frame $(g-i)$ colour are insignificant.

\section{Conclusions}
\label{sec:conclusions}

This study presents an analysis of Milky Way Analogues selected from the SAMI Galaxy Survey \citep{Croom2012,Croom2021} using both the GAMA and Cluster regions to cover a wide range of environmental densities from the field to the most massive clusters. We adopt a nearest neighbours method to obtain close analogues to the Milky Way using four different selection parameters, namely stellar mass $M_{\star}$, star formation rate $SFR$, bulge-to-total ratio $B/T$, and disk effective radius $R_{\rm{e}}$. 

We test which of the 15 combinations of $M_{\star}$, $SFR$, $B/T$, and $R_{\rm{e}}$, impact the MWA selection the most. Our results show that the average properties of the MWAs are most biased if we exclude stellar mass and star formation rate (Fig. \ref{fig:Averages_GAMA_plots}). Out of the four parameters, the bulge-to-total ratio is the least important selection parameter as the average values of $M_{\star}$, $SFR$, and $R_{\rm{e}}$, all lie within $1\sigma$ of the Milky Way regardless of whether $B/T$ is included or not. In contrast, the disk effective radius significantly impacts the MWA selection, as the Milky Way is relatively small compared to other galaxies with similar stellar mass. 

When using all four selection parameters simultaneously, we find that the selected MWAs are biased towards lower stellar mass and larger disk effective radii compared to the Milky Way, despite being the ``closest" MWA based on the average properties, consistent with previous studies \citep[e.g.][]{LN2016,Boardman2020a,Boardman2020b,Pilyugin2023}. We emphasise that including the disk effective radius is the most restrictive parameter choice because the Milky Way is atypically compact compared to other galaxies.

We select the top-10 MWAs from both the GAMA and Cluster regions to investigate their stellar kinematic and environmental properties. These top-10 MWAs correspond to 4\% and 7\% of our selected parent samples in the GAMA and Cluster regions of SAMI Survey, respectively, located in the mass range $10.0<\log(M_{\star}/\rm{M}_{\odot})<11.5$. From a visual morphology classification of these top-10 MWA, we find that $90^{+3}_{-17}\%$ of the GAMA MWA have a central bar structure. This is consistent with the Milky Way which has a central bar, but the result is surprising given that visual morphology was not a selection parameter. 

We can infer the mean global stellar kinematic properties of the Milky Way from the chosen MWAs, revealing that our galaxy exhibits characteristics of a fast rotator, featuring an intrinsically thin disk (with an average intrinsic ellipticity of approximately $0.8$). Additionally, our analysis indicates a pronounced rotational component, as evidenced by the edge-on spin parameter $\lambda_{R,\rm{EO}}$ falling within the range of $0.6-0.7$, and a discernible rise in dispersion towards the galactic centre, as illustrated in Fig. \ref{fig:mapping}. We can infer that the Milky Way is a young, star-forming galaxy (i.e. Fig. \ref{fig:M_SFR_gi_selbySFR} and \ref{fig:M_Age_metallicity_alpha}), with the estimated age, metallicity and $\alpha-$abundance (within 1 $R_{\rm e}$) in the ranges of $1.4\lesssim\rm{Age/Gyr}\lesssim4.6$, $-0.4\lesssim[\rm{Z}/\rm{H}]\lesssim0.2$ and $0.0\lesssim[\alpha/\rm{Fe}]\lesssim0.4$. respectively.

Lastly, we find no significant differences in the properties of the top-10 MWAs from the GAMA and Cluster regions. Both groups cover a wide range in mean environmental densities and have comparable edge-on spin parameter values. This suggests that MWAs can exist in Cluster environments and our selection method successfully picks out MWAs despite residing in clusters. Furthermore, for the top-10 GAMA MWAs, we find that nine galaxies reside in groups with different multiplicities, ranging from pairs to groups with 20 members.

In conclusion, using a nearest neighbours method for selecting MWAs, our work demonstrates the importance of using multiple parameters for comparing the Milky Way to other galaxies, whilst simultaneously offering a new way to find Milky Way-like galaxies when the number of observed galaxies is limited. 

\section*{Acknowledgements}

The SAMI Galaxy Survey is based on observations made at the Anglo-Australian Telescope (AAT). The Sydney-AAO Multi-object Integral ﬁeld spectrograph (SAMI) was developed jointly by the University of Sydney and the Australian Astronomical Observatory. The SAMI input catalogue is based on data taken from the Sloan Digital Sky Survey, the GAMA Survey, and the VST ATLAS Survey. The
SAMI Galaxy Survey website is \url{http://www.sami-survey.org/}. The SAMI Galaxy Survey is supported by the Australian Research Council Centre of Excellence for All Sky Astrophysics in 3 Dimensions (ASTRO 3D), through project number CE170100013, the Australian Research Council Centre of Excellence for All-sky Astrophysics (CAASTRO), through project number CE110001020, and other participating institutions. We acknowledge the traditional owners of the land on which the AAT stands, the Gamilaraay people, and pay our respects to elders past and present. 

GAMA is a joint European–Australasian project based around a spectroscopic campaign using the Anglo-Australian Telescope. The GAMA input catalogue is based on data taken from the SDSS and UKIRT Infrared Deep Sky Survey. Complementary imaging of the GAMA regions is being obtained by a number of independent survey programmes including GALEX MIS, VST-KIDS, VISTA-VIKING, WISE, Herschel-ATLAS, GMRT and ASKAP providing ultraviolet to radio coverage. GAMA is funded by the STFC (UK), the Astrophysical Research Consortium (ARC) (Australia), the AAO, and the Participating Institutions. The
GAMA website is \url{http://www.gama-survey.org/}.

The construction of GSWLC was funded through NASA ADAP award NNX12AE06G. Funding for SDSS-III has been provided by the Alfred P. Sloan Foundation, the Participating Institutions, the National Science Foundation, and the U.S. Department of Energy Ofﬁce of Science. The SDSS-III website is \url{http://www.sdss3.org/}. Based on observations made with the NASA Galaxy Evolution Explorer (GALEX). GALEX is operated for NASA by the California Institute of Technology under NASA contract NAS5-98034. This publication makes use of data products from the Wide-ﬁeld Infrared Survey Explorer, which is a joint project of the University of California, Los Angeles, and the Jet Propulsion Laboratory/California Institute of Technology, funded by the National Aeronautics and Space Administration.

ST acknowledges the support from the Royal Thai Government Scholarship and the University of Sydney Postgraduate Research Supplementary Scholarship in Understanding the Milky Way through Integral Field Spectroscopy. JBH is supported by an ARC Laureate Fellowship FL140100278. The SAMI instrument was funded by Bland-Hawthorn's former Federation Fellowship FF0776384, an ARC LIEF grant LE130100198 (PI Bland-Hawthorn) and funding from the Anglo-Australian Observatory. SB acknowledges funding support from the Australian Research Council through a Future Fellowship (FT140101166). JJB acknowledges the support of an Australian Research Council Future Fellowship (FT180100231). AR acknowledges the receipt of a Scholarship for International Research Fees (SIRF) and an International Living Allowance Scholarship (Ad Hoc Postgraduate Scholarship) at The University of Western Australia. SMS acknowledges funding from the Australian Research Council (DE220100003).

\section*{Data Availability}

The SAMI data presented in this paper are available from Astronomical Optics’ Data Central service at: \url{https://datacentral.org.au/}. The
GAMA data are available at: \url{http://www.gama-survey.org/}. The GSWLC-2 data are available at: \url{https://salims.pages.iu.edu/gswlc/}.


\bibliographystyle{mnras}
\bibliography{Writing}



\appendix

\section{Determining the Milky Way star formation rate (SFR)}
\label{appendix: SFR}

The Milky Way SFR derived by \citet{LN2015} was $1.65\pm0.19$ $\rm{M}_{\odot}\rm{yr}^{-1}$. By assuming a Kroupa IMF and Kroupa-normalised ionising photon rate \citep{Kennicutt1998,KroupaWeidner2003}, this SFR was calculated by re-normalising SFR measurements from the previous literature \citep{Chomiuk2011} and is effectively a variance-weighted mean. However, it is unclear to what degree the SFR measurements are independent. A more conservative uncertainty can be estimated using the variance-weighted standard deviation, rather than the variance-weighted error on the mean. This is also an appropriate method for the bulge mass (See Appendix \ref{appendix: B/T} below), given that there are multiple bulge mass estimates with small uncertainties that are in disagreement with one another. The average SFR of the Milky Way is then
\begin{equation}
    \bar{SFR}=\frac{\sum^{n}_{i=0}w_{i}SFR_{i}}{\sum^{n}_{i=0}w_{i}},
\end{equation}
where $SFR_{i}$ is a SFR from an individual measurement. The weight, $w_{i}$, is given by
\begin{equation}
    w_{i}=\frac{1}{\left(\Delta SFR_{i}\right)^{2}},
\end{equation}
where $\Delta SFR_{i}$ is the uncertainty on an individual measurement. 

The SFRs used for the estimation were from 12 studies which are shown in Table \ref{tab:SFR} \citep{Chomiuk2011,Elia2022}. We derived the uncertainty of our average SFR from the weighted standard deviation, which is given by
\begin{equation}
    \Delta SFR=\sqrt{\frac{\sum^{n}_{i=0}w_{i}\left(SFR_{i}-\bar{SFR}\right)^{2}}{\frac{\left(n-1\right)\sum^{n}_{i=0}w_{i}}{n}}}.
\end{equation}
This results in a derived Milky Way SFR of $1.78\pm0.36$ $\rm{M}_{\odot}\rm{yr}^{-1}$, which is consistent with previous studies \citep{Chomiuk2011,LN2015}.

\begin{table*}
\centering
\small
 \caption{The star formation rates of the Milky Way from previous literature with the methods and uncertainties \citep{Chomiuk2011,Elia2022}.}
 \label{tab:SFR}
 \begin{tabular}{lcl}
\hline
Method & $SFR$ $(\rm{M}_{\odot}\rm{yr}^{-1})$ & Reference  \\
\hline
Ionisation rate from radio free-free emission& $2.0\pm0.6$&\citet{GuestenMezger1982} \\
Ionisation rate from radio free-free emission &$1.6\pm0.5$&\citet{Mezger1987} \\
Ionisation rate from [N II] 205 $\mu$m (COBE)&$2.6\pm1.3$&\citet{Bennett1994} \\
Ionisation rate from [N II] 205 $\mu$m (COBE)&$2.0\pm1.0$&\citet{McKeeWilliam1997} \\
O/B star counts&$1.8\pm0.6$&\citet{Reed2005} \\
Nucleosynthesis from $^{26}$Al (INTEGRAL)&$2.0\pm1.2$&\citet{Diehl2006} \\
Continuum emission at 100 $\mu$m (COBE)&$1.9\pm0.8$&\citet{Misiriotis2006} \\
Ionisation rate from microwave free-free emission (WMAP)&$2.4\pm1.2$&\citet{MurrayRahman2010} \\
YSO counts (Spitzer)&$1.1\pm0.4$&\citet{RobitailleWhitney2010} \\
YSO counts (MSX)&$1.8\pm0.3$&\citet{Davies2011} \\
Continuum emission at 70 $\mu$m (Herschel)&$2.1\pm0.4$&\citet{NoriegaCrespo2013} \\
FIR clump counts (Herschel)&$2.0\pm0.7$&\citet{Elia2022} \\
\hline
  \end{tabular}
\end{table*}

\section{Determining the Milky Way bulge-to-total ratio}
\label{appendix: B/T}

\citet{LN2015,LN2016} published a $B/T$ of $0.16\pm0.03$ where the uncertainty was small and inconsistent with more recent measurements. Therefore, we re-calculated the Milky Way $B/T$ by adding more recent datasets of the bulge stellar mass $M_{\rm{B}}$, scaling the values with the new Galactocentric radius $R_{0}$ and then averaging all values. Note that we define a bulge component as a central component which may include a bar or a pseudo-bulge \citep{Barsanti2021b,Casura2022}.

\begin{figure*}
    \centering
    \small
    \includegraphics[scale=0.83,trim={1.5cm 0.5cm 2.5cm 2.0cm},clip]{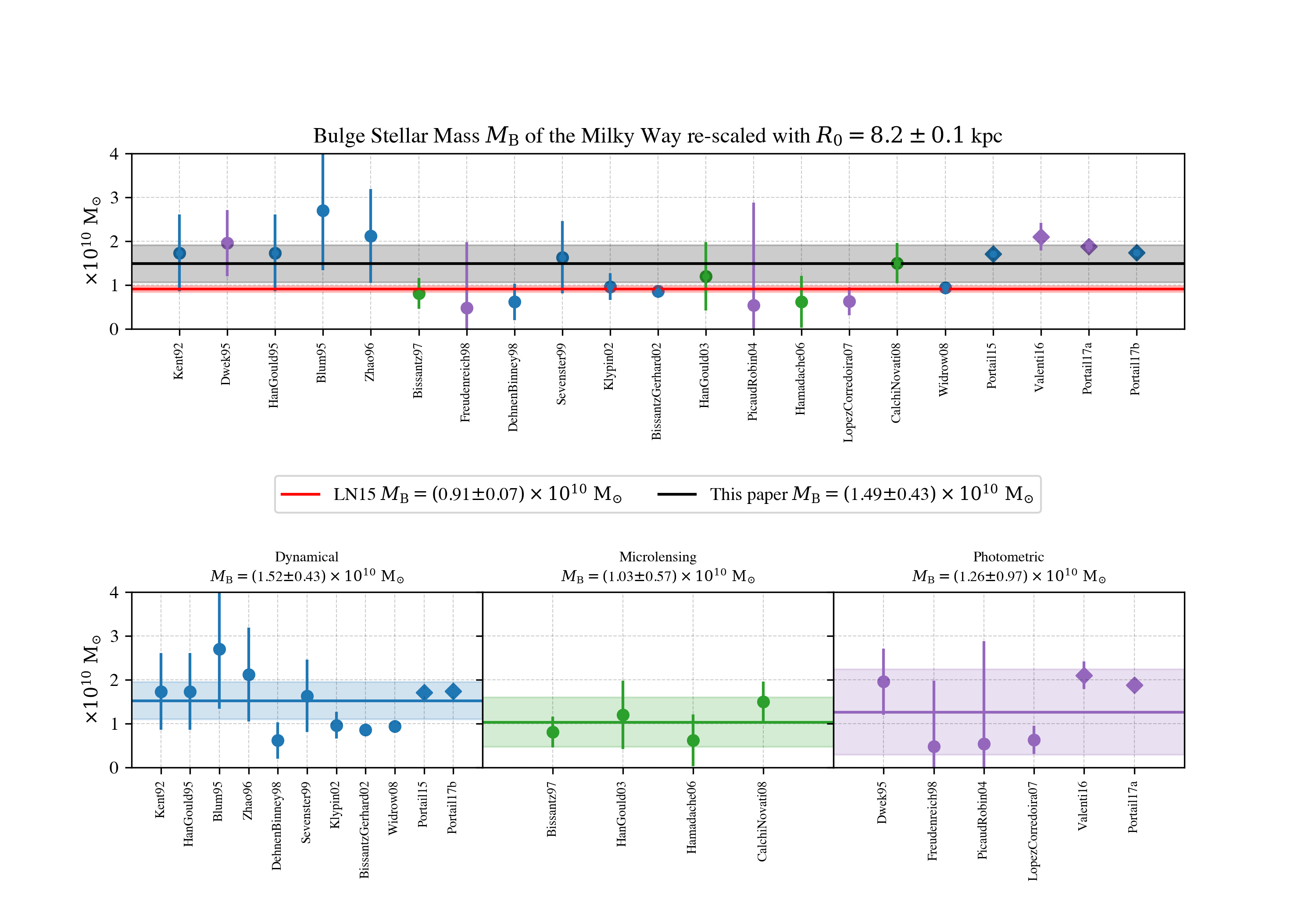}
    \caption{The bulge stellar mass, $M_{\rm{B}}$, of the Milky Way from 21 datasets which was re-scaled with the Galactocentric $R_{0}=8.2\pm0.1$ kpc. The datasets were derived from 3 constraint types including dynamical (blue), microlensing (green) and photometric (purple). The circles are the datasets provided by \citet{LN2015}, while the diamonds are the additional datasets from \citet{Portail2015}, \citet{Valenti2016}, \citet{Portail2017a} and \citet{Portail2017b}.
    The top diagram contains all 21 datasets. The black line is the average $M_{\rm{B}}$ from all constraint types. The red line is the $M_{\rm{B}}$ calculated by the Hierarchical Bayesian Meta-Analysis \citep{LN2015,LN2016}. In the bottom row, the average $M_{\rm{B}}$ from each constraint type is fitted as a solid line and stated on the top of each diagram.}
    \label{fig:MB_plots}
\end{figure*}

There are 21 datasets of $M_{\rm{B}}$ being derived using three different types of measurements including dynamical, photometric and microlensing. We divided these datasets into two groups, the first one including 17 datasets from \citet{LN2015} and another one including 4 datasets from \citet{Portail2015}, \citet{Valenti2016}, \citet{Portail2017a} and \citet{Portail2017b}. The first group was previously scaled by the Galactocentric radius $8.33\pm0.35$ kpc \citep{Gillessen2009} and a normalisation factor of $0.94\pm0.02$ to exclude the contribution from brown dwarfs for the datasets with the dynamical constraint \citep{LN2015}. The second group applied other Galactocentric radii without additional normalisation factors. We re-scaled both groups of datasets to realise the same $R_{0}$ at $8.2\pm0.1$ kpc \citep{BH2016} based on a power-law relationship between $M_{\rm{B}}$ and $R_{0}$. At the same time, we re-scaled the second group with the normalisation factor of $0.94\pm0.02$ for the dynamical-constraint datasets. All re-scaled datasets with their constraint types, power-law relationships and references are listed in Table \ref{tab:M_B} and visualised in Fig. \ref{fig:MB_plots}.

\begin{table*}
\centering
\small
 \caption{The bulge stellar mass of the Milky Way before re-scaling $M_{\rm{B,before}}$ with the associated Galactocentric radius $R_{0,\rm{before}}$ and after re-scaling $M_{\rm{B,after}}$ by the new Galactocentric radius $R_{0}=8.2\pm0.1$ kpc \citep{BH2016} with uncertainties, reference, constraint type, power-law relationship $\beta$ of 21 datasets. A normalisation factor of $0.94\pm0.02$ was applied to \citet{Portail2015} and \citet{Portail2017b} to exclude the contribution from brown dwarfs for the datasets with the dynamical constraint \citep{LN2015}. The power-law relationship $\beta$ is denoted as $M_{\rm{B}}\propto R_{0}^{\beta}$.}
 \label{tab:M_B}
 \begin{tabular}{lccccc}
\hline
Reference & $M_{\rm{B,before}}/\rm{M}_{\odot}$ & $R_{0,\rm{before}}$/kpc & Constraint type & $\beta$ & $M_{\rm{B,after}}/\rm{M}_{\odot}$  \\
\hline
\citet{Kent1992}&1.76 $\pm$ 0.88& 8.33 $\pm$ 0.35 &Dynamical&1&1.73 $\pm$ 0.87 \\
\citet{Dwek1995}&2.02 $\pm$ 0.78& 8.33 $\pm$ 0.35 &Photometric&2&1.96 $\pm$ 0.76 \\
\citet{HanandGould1995}&1.76 $\pm$ 0.88& 8.33 $\pm$ 0.35 &Dynamical&1&1.73 $\pm$ 0.87 \\
\citet{Blum1995}&2.74 $\pm$ 1.37& 8.33 $\pm$ 0.35 &Dynamical&1&2.70 $\pm$ 1.36 \\
\citet{Zhao1996}&2.15 $\pm$ 1.08& 8.33 $\pm$ 0.35 &Dynamical&1&2.12 $\pm$ 1.07 \\
\citet{Bissant1997}&0.81 $\pm$ 0.22& 8.33 $\pm$ 0.35 &Microlensing&0&0.81 $\pm$ 0.35 \\
\citet{Freudenreich1998}&0.48 $\pm$ 0.65& 8.33 $\pm$ 0.35 &Photometric&0&0.48 $\pm$ 1.50 \\
\citet{DehnenandBinney1998}&0.62 $\pm$ 0.38& 8.33 $\pm$ 0.35 &Dynamical&0.5&0.62 $\pm$ 0.41 \\
\citet{Sevenster1999}&1.66 $\pm$ 0.83& 8.33 $\pm$ 0.35 &Dynamical&1&1.63 $\pm$ 0.82 \\
\citet{Klypin2002}&0.98 $\pm$ 0.31& 8.33 $\pm$ 0.35 &Dynamical&1&0.96 $\pm$ 0.31 \\
\citet{BissantzandGerhard2002}&0.87 $\pm$ 0.09& 8.33 $\pm$ 0.35 &Dynamical&1&0.86 $\pm$ 0.09 \\
\citet{HanandGould2003}&1.20 $\pm$ 0.60& 8.33 $\pm$ 0.35 &Microlensing&0&1.20 $\pm$ 0.78 \\
\citet{Picaud2004}&0.54 $\pm$ 1.11& 8.33 $\pm$ 0.35 &Photometric&0&0.54 $\pm$ 2.34 \\
\citet{Hamadache2006}&0.62 $\pm$ 0.31& 8.33 $\pm$ 0.35 &Microlensing&0&0.62 $\pm$ 0.59 \\
\citet{LopezCorredoira2007}&0.65 $\pm$ 0.33& 8.33 $\pm$ 0.35 &Photometric&2&0.63 $\pm$ 0.32 \\
\citet{CalchiNovati2008}&1.50 $\pm$ 0.38& 8.33 $\pm$ 0.35 &Microlensing&0&1.50 $\pm$ 0.46 \\
\citet{Widrow2008}&0.95 $\pm$ 0.12& 8.33 $\pm$ 0.35 &Dynamical&1&0.94 $\pm$ 0.12 \\
\citet{Portail2015}&1.84 $\pm$ 0.07&  8.30 $\pm$ 0.00  &Dynamical&1&1.70 $\pm$ 0.07 \\
\citet{Valenti2016}&2.00 $\pm$ 0.30&  8.00 $\pm$ 0.00  &Photometric&2&2.05 $\pm$ 0.30 \\
\citet{Portail2017a}&1.88 $\pm$ 0.12&  8.20 $\pm$ 0.10  &Photometric&2&1.88 $\pm$ 0.12 \\
\citet{Portail2017b}&1.85 $\pm$ 0.05&  8.20 $\pm$ 0.10  &Dynamical&1&1.74 $\pm$ 0.06 \\
\hline
  \end{tabular}
\end{table*}

We calculated the averaged bulge stellar mass and its uncertainty using the same method for the Milky Way SFR in Appendix \ref{appendix: SFR}. Such an approach leads to a larger uncertainty than \citet{LN2015}, but is reasonable given that multiple individual mass estimates have small uncertainties, but are in disagreement with each other. The derived Milky Way bulge stellar mass is $(1.49\pm0.43)\times10^{10}~\rm{M}_{\odot}$.

Finally, we calculated the Milky Way $B/T$ from the ratio between the bulge stellar mass and the total stellar mass. We applied the total stellar mass at $(5\pm1)\times10^{10} ~\rm{M}_{\odot}$ \citep{BH2016} and then obtained the Milky Way $B/T$ at $0.30\pm0.10$ which is consistent with previous studies \citep{BH2016,Barbuy2018}.

\section{The averaged properties of the MWAs as a function of the number of the galaxies being selected}
\label{appendix:bias}

\begin{figure*}
    \centering
    \small
    \includegraphics[scale=0.6,trim={0cm 0cm 0cm 0cm},clip]{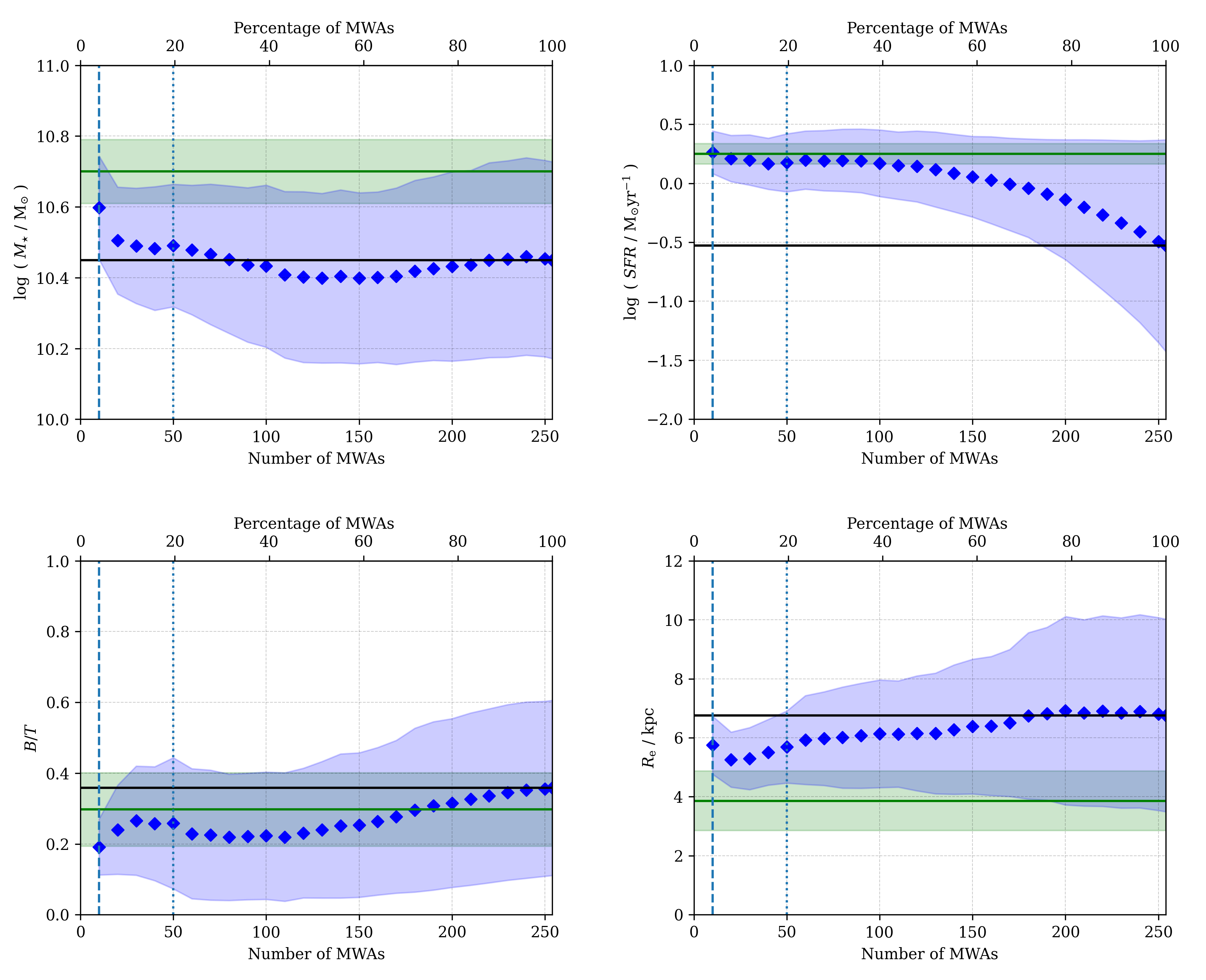}
    \caption{The averaged properties of selected MWAs as a function of the number of such MWAs in the GAMA regions. Here we indicate four properties $\log(M_{\star})$, $\log(SFR)$, $B/T$ and $R_{\rm{e}}$. The green solid lines and shades represent the Milky Way's values and their $1\sigma$ uncertainties. The black solid lines show the averages of the sample. The light blue dash and dotted lines mark the top-10 and top-50 MWAs, respectively. The blue diamonds and shades illustrate the averages and their standard deviations with an interval of 10 between each diamond (except the last interval as the last marker equals the number of the entire sample). The diagrams show that the averages move towards the averages of the sample when we increase the number of selected MWAs and hence represent the sample rather than Milky Way-like galaxies.}
    \label{fig:Distance_GAMA}
\end{figure*}

\begin{figure*}
    \centering
    \small
    \includegraphics[scale=0.6,trim={0cm 0cm 0cm 0cm},clip]{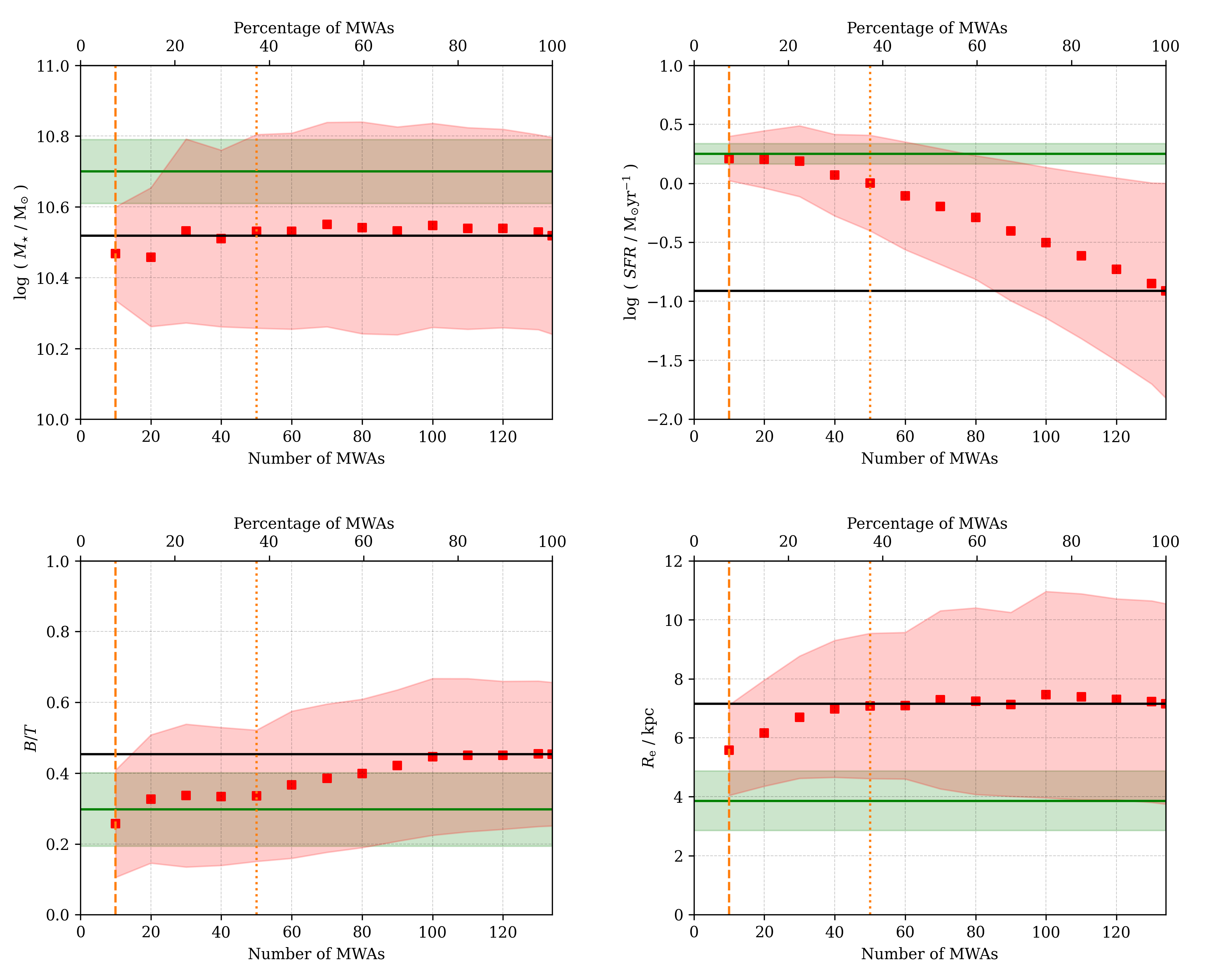}
    \caption{The averaged properties of selected MWAs as a function of the number of such MWAs in the Cluster regions. Here we indicate four properties $\log(M_{\star})$, $\log(SFR)$, $B/T$ and $R_{\rm{e}}$. The green solid lines and shades represent the Milky Way's values and their $1\sigma$ uncertainties. The black solid lines show the averages of the sample. The orange dash and dotted lines mark the top-10 and top-50 MWAs, respectively. The red squares and shades illustrate the averages and their standard deviations with an interval of 10 between each square (except the last interval as the last marker equals the number of the entire sample). The diagrams show that the averages move towards the averages of the sample when we increase the number of selected MWAs and hence represent the sample rather than Milky Way-like galaxies.}
    \label{fig:Distance_Cluster}
\end{figure*}

In this section, we determine four properties of the MWAs as a function of the number of galaxies being selected as MWAs including stellar mass $\log(M_{\star})$, star formation rate $\log(SFR)$, bulge-to-total ratio $B/T$ and disk effective radius $R_{\rm{e}}$. We calculate the distance $D$ from the Milky Way's values for the entire samples using Eq. \ref{eq:nearest_neighbours}. We selected the first 10 galaxies with the closest distance $D$ to the Milky Way as the MWAs and calculated their averaged properties. Then we increased the number of selected MWAs by a step of 10 and repeated the calculation until the number of selected MWAs equals the number of galaxies in the entire samples. We showed the averaged values and their standard deviations, the Milky Way's values and the averaged values of the entire samples in Fig. \ref{fig:Distance_GAMA} and \ref{fig:Distance_Cluster} for the GAMA and Cluster regions, respectively. 

When we increase the number of MWAs, the averaged properties of the MWAs become closer to the averaged properties of the entire samples. At the same time, the standard deviations also grow when increasing the number of MWAs. There are three special cases to be further explained. The first case is the averaged $B/T$ in the GAMA regions: at the low number of MWAs, the averaged values move towards the Milky Way value, but when the number of MWAs exceeds 180 the averaged values finally move away from the Milky Way to the entire sample. This result is probably driven by including passive galaxies with high $B/T$ as fraction gets above $\sim60\%$. The second case is the averaged $\log(M_{\star})$ in both regions where the changes are marginal, i.e. the averaged values locate nearby the averages of the samples regardless of the number of MWAs. This result is due to to the bias in the stellar mass function, i.e., most of the galaxies in the samples have lower stellar masses than the Milky Way. The last case is the averaged $R_{\rm{e}}$ in both regions whose the locations are greater than $1\sigma$ from the Milky Way's $R_{\rm{e}}$ at $3.86\pm1.01$ kpc for all cases. This result is caused by the bias in the disk effective radius, as the Milky Way is relatively small compared to other galaxies with similar stellar mass.

Unsurprisingly, when we increase the number of galaxies being selected as MWAs, the averaged properties represent the entire sample instead of Milky Way-like galaxies. Therefore, we prefer to select the top-10 galaxies as MWAs rather than applying a larger number to the selection.

\section{Calculation of MWA properties based on 15 different parameter combinations}
\label{appendix:MWAs_prop}

After analysing 15 parameter combinations, we select 15 sets of top-10 MWAs and average their quantitative properties including stellar mass $\log(M_{\star})$, star formation rate $\log(SFR)$, bulge-to-total ratio $B/T$, disk effective radius $R_{\rm{e}}$ in kpc, edge-on spin parameter $\lambda_{R,\rm{EO}}$, age in Gyr, metallicity [Z/H], $\alpha-$abundance [$\alpha$/Fe] and rest-frame $(g-i)$ colour. We calculate the standard deviation as the uncertainty for each parameter in each selection combination.

\begin{figure*}
    \centering
    \small
    \includegraphics[scale=0.95,trim={0cm 0cm 0cm 0cm},clip]{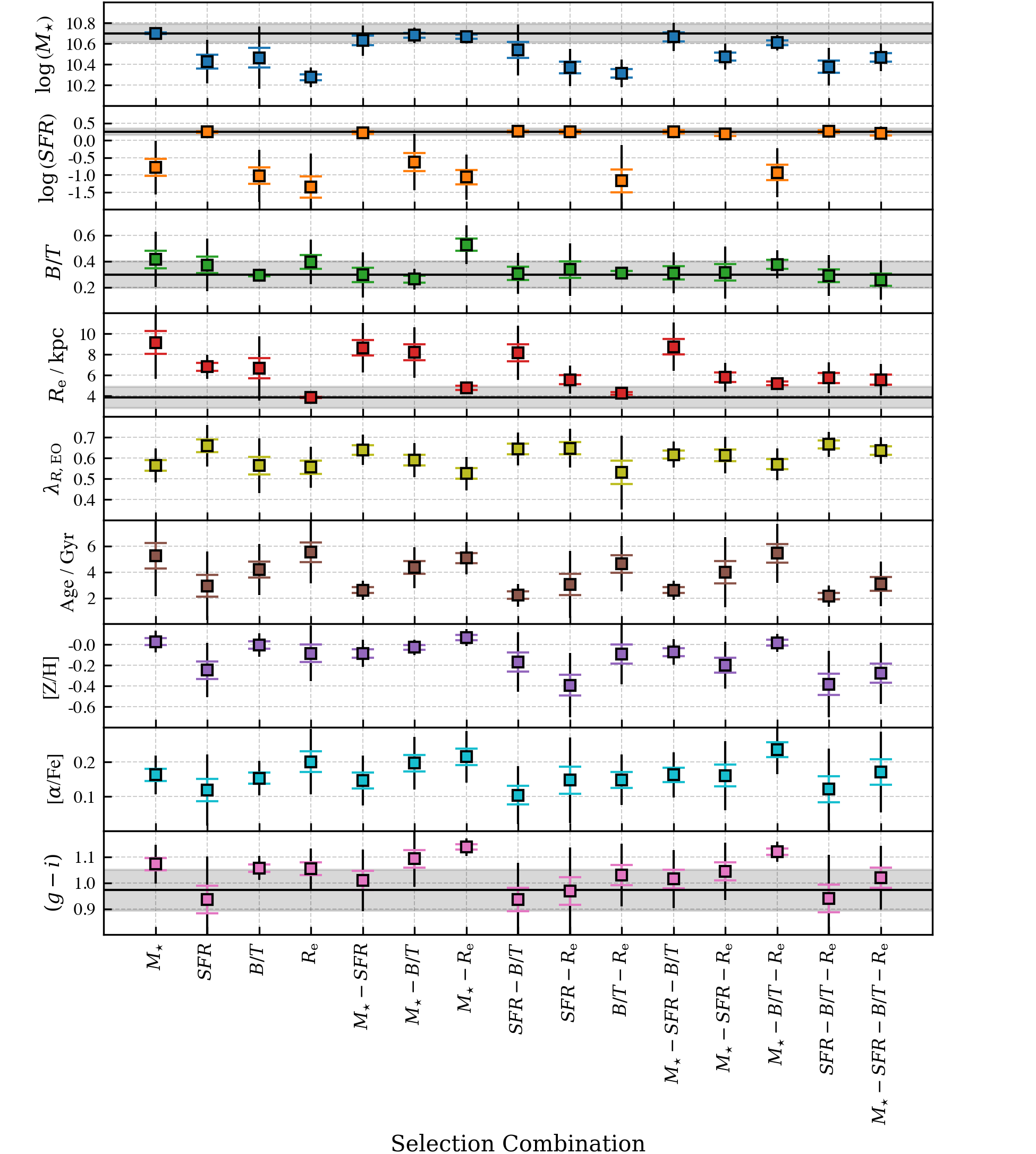}
    \caption{Comparisons between the averaged properties of the top-10 Cluster MWAs selected by 15 parameter combinations. There are nine properties from top to bottom including $\log(M_{\star})$ (blue), $\log(SFR)$ (orange), $B/T$ (green), $R_{\rm{e}}$ (red), $\lambda_{R,\rm{EO}}$ (olive), age (brown), metallicity $[\rm{Z}/\rm{H}]$ (purple), $\alpha-$abundance $[\alpha/\rm{Fe}]$ (cyan) and rest-frame colour $(g-i)$ (pink). All square markers are accompanied by standard deviations (black error bars) and error on the mean (coloured caps). The Milky Way parameter in each diagram is shown as a black solid line with $1\sigma$ grey shaded area except $\lambda_{R,\rm{EO}}$, age, $[\rm{Z}/\rm{H}]$ and $[\alpha/\rm{Fe}]$. These comparisons show that each parameter combination provides MWAs with different properties. We find that some averaged properties are inconsistent with those of the Milky Way when this specific parameter is excluded from the selection combination.}
    \label{fig:Averages_Cluster_plots}
\end{figure*}

We illustrate the averaged properties of the selected MWAs separately for the GAMA and Cluster regions in Table \ref{tab:GAMA_MWAs} and \ref{tab:cluster_MWAs}, respectively. We include the Milky Way parameters within both tables for comparison. Note that the Milky Way $\lambda_{R,\rm{EO}}$, age, [Z/H] and [$\alpha$/Fe] are unknown. The results are presented in Fig. \ref{fig:Averages_GAMA_plots} and Fig. \ref{fig:Averages_Cluster_plots} for the GAMA and Cluster MWAs, respectively.

\begin{table*}
\centering
\smaller
 \caption{The averaged properties of the GAMA MWAs selected by 15 different parameter combinations with uncertainties.}
 \label{tab:GAMA_MWAs}
 \begin{tabular}{lccccccccc}
  \hline
  \textbf{Selection combination}&$\log$($M_{\star}$)&$\log$($SFR$)&$B/T$&$R_{\rm{e}}$ / $\rm{kpc}$&$\lambda_{R,\rm{EO}}$&Age / Gyr&[Z/H]&[$\alpha$/Fe]&$(g-i)$\\
    \hline
$M_{\star}$&10.70 $\pm$ 0.02&-0.66 $\pm$ 1.01&0.32 $\pm$ 0.24&7.42 $\pm$ 1.70&0.53 $\pm$ 0.11&4.75 $\pm$ 2.80&-0.09 $\pm$ 0.15&0.16 $\pm$ 0.06&1.09 $\pm$ 0.11\\
$SFR$&10.36 $\pm$ 0.15&0.26 $\pm$ 0.02&0.31 $\pm$ 0.29&10.44 $\pm$ 4.67&0.64 $\pm$ 0.11&2.10 $\pm$ 0.77&-0.31 $\pm$ 0.24&0.11 $\pm$ 0.11&0.88 $\pm$ 0.08\\
$B/T$&10.38 $\pm$ 0.24&-0.62 $\pm$ 0.91&0.29 $\pm$ 0.01&5.74 $\pm$ 2.32&0.56 $\pm$ 0.12&3.53 $\pm$ 1.27&-0.19 $\pm$ 0.32&0.18 $\pm$ 0.08&1.00 $\pm$ 0.12\\
$R_{\rm{e}}$&10.26 $\pm$ 0.15&-0.60 $\pm$ 0.89&0.35 $\pm$ 0.18&3.89 $\pm$ 0.06&0.42 $\pm$ 0.18&3.92 $\pm$ 2.42&-0.36 $\pm$ 0.37&0.14 $\pm$ 0.11&1.01 $\pm$ 0.09\\
$M_{\star}-SFR$&10.63 $\pm$ 0.12&0.27 $\pm$ 0.14&0.11 $\pm$ 0.08&7.91 $\pm$ 2.34&0.69 $\pm$ 0.09&2.02 $\pm$ 0.64&-0.02 $\pm$ 0.16&0.10 $\pm$ 0.10&0.93 $\pm$ 0.09\\
$M_{\star}-B/T$&10.71 $\pm$ 0.05&-0.66 $\pm$ 0.83&0.30 $\pm$ 0.07&8.15 $\pm$ 2.56&0.53 $\pm$ 0.12&4.84 $\pm$ 2.21&-0.01 $\pm$ 0.11&0.20 $\pm$ 0.07&1.11 $\pm$ 0.08\\
$M_{\star}-R_{\rm{e}}$&10.63 $\pm$ 0.06&-0.70 $\pm$ 1.00&0.50 $\pm$ 0.29&4.28 $\pm$ 0.72&0.57 $\pm$ 0.19&4.79 $\pm$ 3.12&-0.02 $\pm$ 0.16&0.22 $\pm$ 0.10&1.06 $\pm$ 0.10\\
$SFR-B/T$&10.38 $\pm$ 0.33&0.24 $\pm$ 0.08&0.30 $\pm$ 0.10&5.68 $\pm$ 1.80&0.65 $\pm$ 0.14&2.17 $\pm$ 0.66&-0.35 $\pm$ 0.27&0.15 $\pm$ 0.08&0.89 $\pm$ 0.08\\
$SFR-R_{\rm{e}}$&10.20 $\pm$ 0.13&0.24 $\pm$ 0.08&0.32 $\pm$ 0.18&4.49 $\pm$ 0.84&0.62 $\pm$ 0.17&2.01 $\pm$ 0.63&-0.44 $\pm$ 0.32&0.14 $\pm$ 0.07&0.88 $\pm$ 0.08\\
$B/T-R_{\rm{e}}$&10.34 $\pm$ 0.13&-1.24 $\pm$ 1.06&0.28 $\pm$ 0.02&3.79 $\pm$ 0.44&0.55 $\pm$ 0.15&4.25 $\pm$ 1.98&-0.04 $\pm$ 0.09&0.15 $\pm$ 0.08&1.08 $\pm$ 0.06\\
$M_{\star}-SFR-B/T$&10.65 $\pm$ 0.13&0.26 $\pm$ 0.16&0.15 $\pm$ 0.09&7.35 $\pm$ 1.83&0.67 $\pm$ 0.08&2.07 $\pm$ 0.68&-0.05 $\pm$ 0.16&0.15 $\pm$ 0.09&0.96 $\pm$ 0.07\\
$M_{\star}-SFR-R_{\rm{e}}$&10.54 $\pm$ 0.15&0.28 $\pm$ 0.12&0.31 $\pm$ 0.24&5.80 $\pm$ 0.81&0.61 $\pm$ 0.14&2.02 $\pm$ 0.71&-0.05 $\pm$ 0.14&0.14 $\pm$ 0.08&0.96 $\pm$ 0.07\\
$M_{\star}-B/T-R_{\rm{e}}$&10.64 $\pm$ 0.09&-0.10 $\pm$ 0.69&0.24 $\pm$ 0.09&5.06 $\pm$ 0.78&0.59 $\pm$ 0.17&4.24 $\pm$ 2.67&-0.05 $\pm$ 0.15&0.21 $\pm$ 0.08&1.04 $\pm$ 0.10\\
$SFR-B/T-R_{\rm{e}}$&10.20 $\pm$ 0.13&0.22 $\pm$ 0.10&0.30 $\pm$ 0.11&4.36 $\pm$ 0.82&0.56 $\pm$ 0.22&2.02 $\pm$ 0.63&-0.47 $\pm$ 0.29&0.16 $\pm$ 0.07&0.91 $\pm$ 0.11\\
$M_{\star}-SFR-B/T-R_{\rm{e}}$&10.60 $\pm$ 0.15&0.26 $\pm$ 0.18&0.19 $\pm$ 0.08&5.75 $\pm$ 0.99&0.65 $\pm$ 0.08&2.22 $\pm$ 0.94&-0.06 $\pm$ 0.15&0.18 $\pm$ 0.11&0.99 $\pm$ 0.10\\
    \hline
  \textbf{Milky Way}&\textbf{10.70 $\pm$ 0.09}&\textbf{0.25 $\pm$ 0.09}&\textbf{0.30 $\pm$ 0.10}&\textbf{3.86 $\pm$ 1.01}&&&&&\textbf{0.972$\pm$0.080} \\
    \hline
 \end{tabular}
\end{table*}

\begin{table*}
\centering
\smaller
 \caption{The averaged properties of the Cluster MWAs selected by 15 different parameter combinations with uncertainties.}
 \label{tab:cluster_MWAs}
 \begin{tabular}{lccccccccc}
  \hline
  \textbf{Selection combination}&$\log$($M_{\star}$)&$\log$($SFR$)&$B/T$&$R_{\rm{e}}$ / $\rm{kpc}$&$\lambda_{R,\rm{EO}}$&Age / Gyr&[Z/H]&[$\alpha$/Fe]&$(g-i)$ \\
  \hline
$M_{\star}$&10.70 $\pm$ 0.02&-0.78 $\pm$ 0.78&0.41 $\pm$ 0.21&9.17 $\pm$ 3.52&0.57 $\pm$ 0.08&5.27 $\pm$ 3.14&0.03 $\pm$ 0.11&0.16 $\pm$ 0.06&1.07 $\pm$ 0.07\\
$SFR$&10.43 $\pm$ 0.21&0.24 $\pm$ 0.06&0.37 $\pm$ 0.20&6.82 $\pm$ 1.15&0.66 $\pm$ 0.10&2.94 $\pm$ 2.63&-0.25 $\pm$ 0.26&0.12 $\pm$ 0.10&0.94 $\pm$ 0.17\\
$B/T$&10.46 $\pm$ 0.30&-1.02 $\pm$ 0.75&0.29 $\pm$ 0.02&6.67 $\pm$ 3.09&0.56 $\pm$ 0.13&4.19 $\pm$ 1.98&-0.00 $\pm$ 0.11&0.15 $\pm$ 0.05&1.06 $\pm$ 0.05\\
$R_{\rm{e}}$&10.28 $\pm$ 0.10&-1.34 $\pm$ 0.97&0.40 $\pm$ 0.17&3.87 $\pm$ 0.14&0.56 $\pm$ 0.10&5.54 $\pm$ 2.42&-0.08 $\pm$ 0.27&0.20 $\pm$ 0.09&1.06 $\pm$ 0.08\\
$M_{\star}-SFR$&10.63 $\pm$ 0.15&0.22 $\pm$ 0.11&0.30 $\pm$ 0.17&8.65 $\pm$ 2.37&0.64 $\pm$ 0.07&2.61 $\pm$ 0.74&-0.08 $\pm$ 0.13&0.15 $\pm$ 0.07&1.01 $\pm$ 0.12\\
$M_{\star}-B/T$&10.68 $\pm$ 0.07&-0.62 $\pm$ 0.81&0.26 $\pm$ 0.08&8.20 $\pm$ 2.43&0.59 $\pm$ 0.08&4.35 $\pm$ 1.58&-0.02 $\pm$ 0.07&0.20 $\pm$ 0.08&1.09 $\pm$ 0.11\\
$M_{\star}-R_{\rm{e}}$&10.67 $\pm$ 0.06&-1.06 $\pm$ 0.66&0.53 $\pm$ 0.15&4.79 $\pm$ 0.58&0.53 $\pm$ 0.08&5.08 $\pm$ 1.25&0.07 $\pm$ 0.08&0.22 $\pm$ 0.07&1.14 $\pm$ 0.03\\
$SFR-B/T$&10.54 $\pm$ 0.25&0.26 $\pm$ 0.08&0.31 $\pm$ 0.16&8.17 $\pm$ 2.63&0.64 $\pm$ 0.08&2.22 $\pm$ 0.88&-0.17 $\pm$ 0.29&0.10 $\pm$ 0.08&0.94 $\pm$ 0.14\\
$SFR-R_{\rm{e}}$&10.37 $\pm$ 0.18&0.25 $\pm$ 0.14&0.34 $\pm$ 0.20&5.58 $\pm$ 1.35&0.65 $\pm$ 0.09&3.05 $\pm$ 2.59&-0.39 $\pm$ 0.31&0.15 $\pm$ 0.12&0.97 $\pm$ 0.17\\
$B/T-R_{\rm{e}}$&10.31 $\pm$ 0.13&-1.17 $\pm$ 1.04&0.31 $\pm$ 0.04&4.26 $\pm$ 0.43&0.53 $\pm$ 0.18&4.64 $\pm$ 2.14&-0.09 $\pm$ 0.29&0.15 $\pm$ 0.07&1.03 $\pm$ 0.12\\
$M_{\star}-SFR-B/T$&10.67 $\pm$ 0.13&0.25 $\pm$ 0.15&0.31 $\pm$ 0.16&8.76 $\pm$ 2.34&0.62 $\pm$ 0.06&2.61 $\pm$ 0.73&-0.07 $\pm$ 0.12&0.16 $\pm$ 0.07&1.02 $\pm$ 0.11\\
$M_{\star}-SFR-R_{\rm{e}}$&10.48 $\pm$ 0.12&0.18 $\pm$ 0.16&0.31 $\pm$ 0.20&5.82 $\pm$ 1.40&0.61 $\pm$ 0.09&4.00 $\pm$ 2.70&-0.20 $\pm$ 0.23&0.16 $\pm$ 0.10&1.05 $\pm$ 0.11\\
$M_{\star}-B/T-R_{\rm{e}}$&10.61 $\pm$ 0.07&-0.93 $\pm$ 0.70&0.38 $\pm$ 0.11&5.22 $\pm$ 0.52&0.57 $\pm$ 0.08&5.46 $\pm$ 2.28&0.02 $\pm$ 0.09&0.24 $\pm$ 0.07&1.12 $\pm$ 0.04\\
$SFR-B/T-R_{\rm{e}}$&10.38 $\pm$ 0.18&0.26 $\pm$ 0.14&0.29 $\pm$ 0.16&5.74 $\pm$ 1.49&0.67 $\pm$ 0.06&2.14 $\pm$ 0.82&-0.38 $\pm$ 0.32&0.12 $\pm$ 0.12&0.94 $\pm$ 0.17\\
$M_{\star}-SFR-B/T-R_{\rm{e}}$&10.47 $\pm$ 0.13&0.21 $\pm$ 0.19&0.26 $\pm$ 0.15&5.58 $\pm$ 1.53&0.64 $\pm$ 0.06&3.10 $\pm$ 1.72&-0.28 $\pm$ 0.30&0.17 $\pm$ 0.12&1.02 $\pm$ 0.12\\
  \hline
  \textbf{Milky Way}&\textbf{10.70 $\pm$ 0.09}&\textbf{0.25 $\pm$ 0.09}&\textbf{0.30 $\pm$ 0.10}&\textbf{3.86 $\pm$ 1.01}&&&&&\textbf{0.972$\pm$0.080}\\
  \hline
 \end{tabular}
\end{table*}

\section{MWA Selection using rest-frame (g-i) colour}
\label{appendix:(g-i) colour}

\begin{figure*}
    \centering
    \includegraphics[scale=0.38,trim={0.2cm 0cm 2.0cm 0cm},clip]{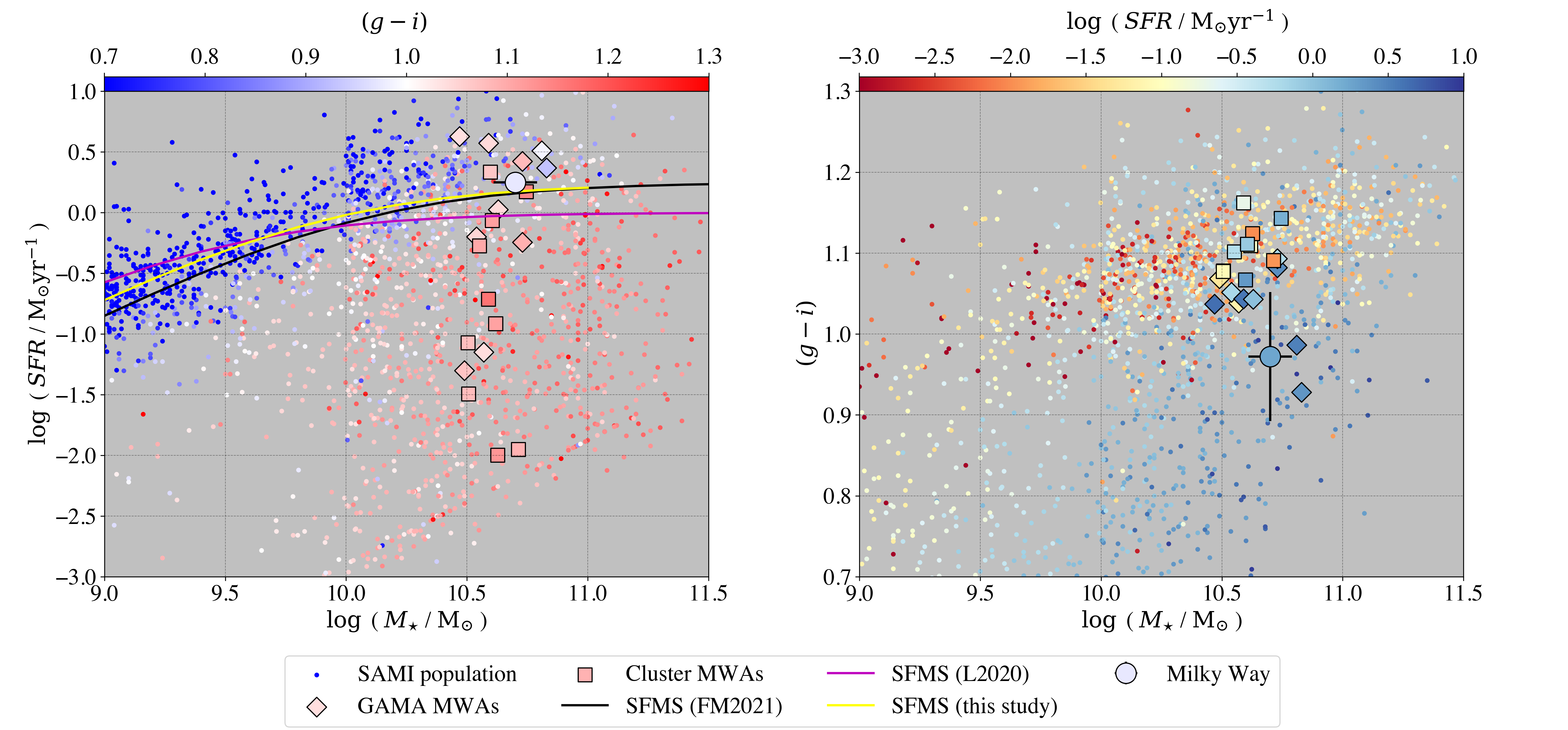}
    \caption{The correlation between the stellar mass $\log(M_{\star})$, the star formation rate $\log(SFR)$ and the rest-frame $(g-i)$ colour of the MWAs selected by stellar mass $\log(M_{\star})$, rest-frame $(g-i)$ colour, bulge-to-total ratio $B/T$ and disk effective radius $R_{\rm{e}}$. All SAMI galaxies are shown as small dots. The MWAs are shown as diamonds for the GAMA MWAs and squares for the Cluster MWAs. The Milky Way is a large dot with $1\sigma$ error bars. \textit{(a) Left panel:} the star formation rates $\log(SFR)$ as a function of stellar mass $\log(M_{\star})$ colour coded by the rest-frame $(g-i)$ colour. The yellow solid curve illustrates the star-forming main sequence line (SFMS) calculated using Eq. \ref{eq:SFMS}. The black solid curve illustrates the SFMS calculated by \citet{FraserMcKelvie2021} which is consistent with the calculation in this paper. The magenta solid curve illustrates the SFMS calculated by \citet{Leslie2020} which is different from the calculation in this paper due to different sample. \textit{(b) Right panel:} the rest-frame $(g-i)$ colour as a function of stellar mass $\log(M_{\star})$ colour coded by the star formation rates $\log(SFR)$. This figure illustrates that the rest-frame $(g-i)$ colour of the MWAs are distributed within $3\sigma$ from the Milky Way at $0.972 \pm 0.080$ \citep{LNB2015,LN2016}. However, the SFRs of the MWAs scatter between $-2.0>\log(SFR/\rm{M}_{\odot}\rm{yr}^{-1})>1.0$.}
    \label{fig:M_SFR_gi_selbygi}
\end{figure*}

Here we consider selecting the top-10 MWAs based on four selection parameters including stellar mass $\log(M_{\star})$, rest-frame $(g-i)$ colour, bulge-to-total ratio $B/T$ and disk effective radius $R_{\rm{e}}$.

The observed $(g-i)$ colour of the SAMI galaxies was derived by photometric fitting from the GAMA catalogue \citep{Hill2011,Liske2015}, the Ninth Data Release of the Sloan Digital Sky Survey (SDSS DR9) imaging data \citep{Ahn2012} and the VLT Survey Telescope (VST) ATLAS imaging data \citep{Shanks2013}. The observed $(g-i)$ colours provided by \code{InputCatGAMADR3} and \code{InputClusterDR3} catalogues were previously corrected for Galactic extinction \citep{Bryant2015,Owers2017} but not $k-$corrected. Hence, we derived the rest-frame $(g-i)$ colour of the SAMI galaxies so that we could compare them to the Milky Way in terms of rest-frame $(g-i)$ colour. 

For the SAMI population in the GAMA regions, we obtained the $k-$corrections from the catalogue \code{Eqkcorr\_model\_z00v05} \citep{Loveday2012}. The typical $k-$corrections over the low redshift range of the GAMA population were approximately $0.05$ magnitudes. To derive equivalent $k-$corrections for the Cluster population, we used the variation of the $k-$correction in the GAMA population as a function of observed $(g-i)$ colour to estimate a relation that was appropriate across the redshift and colour range of the SAMI population and was consistent with that applied in the GAMA regions. The fitted relation was given by
\begin{equation}
    k = (g-i)_{z}-(g-i)_{0} = 1.66 \times z \times (g-i)_{z},
	\label{eq:k-correction}
\end{equation}
where $k$ is the $k-$correction, $(g-i)_{z}$ is the observed $(g-i)$ colour at a particular redshift $z$ and $(g-i)_{0}$ is the rest-frame $(g-i)$ colour. This estimation provided the rest-frame $(g-i)$ colour with a median of absolute scatters of $0.007$ magnitudes from the values in the catalogue \code{Eqkcorr\_model\_z00v05}.

The Milky Way rest-frame $(g-i)$ colour and its uncertainty were derived from galaxies with the total stellar masses and the SFRs that matched the Milky Way \citep{LNB2015,LN2016}. At $z=0$, the estimated Milky Way rest-frame $(g-r)$ and $(r-i)$ colours are $0.678_{-0.057}^{+0.069}$ and $0.294_{-0.046}^{+0.052}$, respectively. We therefore used the rest-frame $(g-i)$ colour of the Milky Way at $0.972 \pm 0.080$ as a selection parameter. 

The top-10 MWAs from both regions selected by this parameter combination have averaged stellar masses of $10.64\pm0.12$ and $10.61\pm0.07$ for the GAMA and Cluster regions. The averaged rest-frame $(g-i)$ colours are $1.04\pm0.04$ and $1.11\pm0.03$. These averages agree with \citet{LNB2015,LN2016} estimated value of the Milky Way within $1\sigma$. We highlight the top-10 MWAs selected by $\log(M_{\star})-(g-i)-B/T-R_{\rm{e}}$ in Fig. \ref{fig:M_SFR_gi_selbygi}. The MWAs also have similar morphologies and stellar kinematics. However, the averaged $\log(SFRs)$ are significantly lower than that of the Milky Way with large standard deviations: $-0.04\pm0.66$ and $-0.80\pm0.80$ in the GAMA and Cluster MWAs, respectively. This result is caused by the fact that the majority of the MWAs have the SFRs below the SFMS lower boundary as shown in Fig. \ref{fig:M_SFR_gi_selbygi}(a). Hence, this selection results in MWAs with lower SFRs as compared to the Milky Way.

In summary, we caution against substituting the SFR with the rest-frame $(g-i)$ colourbecause these selected MWAs would have the SFRs inconsistent with the Milky Way. Unlike the successful MWA selections with other colours such as $(NUV-r)$ \citep{Krishnarao2020b}, the measured rest-frame $(g-i)$ colour is influenced by several sources including stellar population age, dust attenuation and redshift. Therefore, the relation between SFR and the rest-frame $(g-i)$ colour is not perfectly one-to-one and has a considerable amount of scatter.

\section{The companions of the central and satellite GAMA MWAs}
\label{appendix: companions}

\begin{figure*}
    \centering
    \includegraphics[scale=0.95,trim={0cm 0cm 1.0cm 0cm},clip]{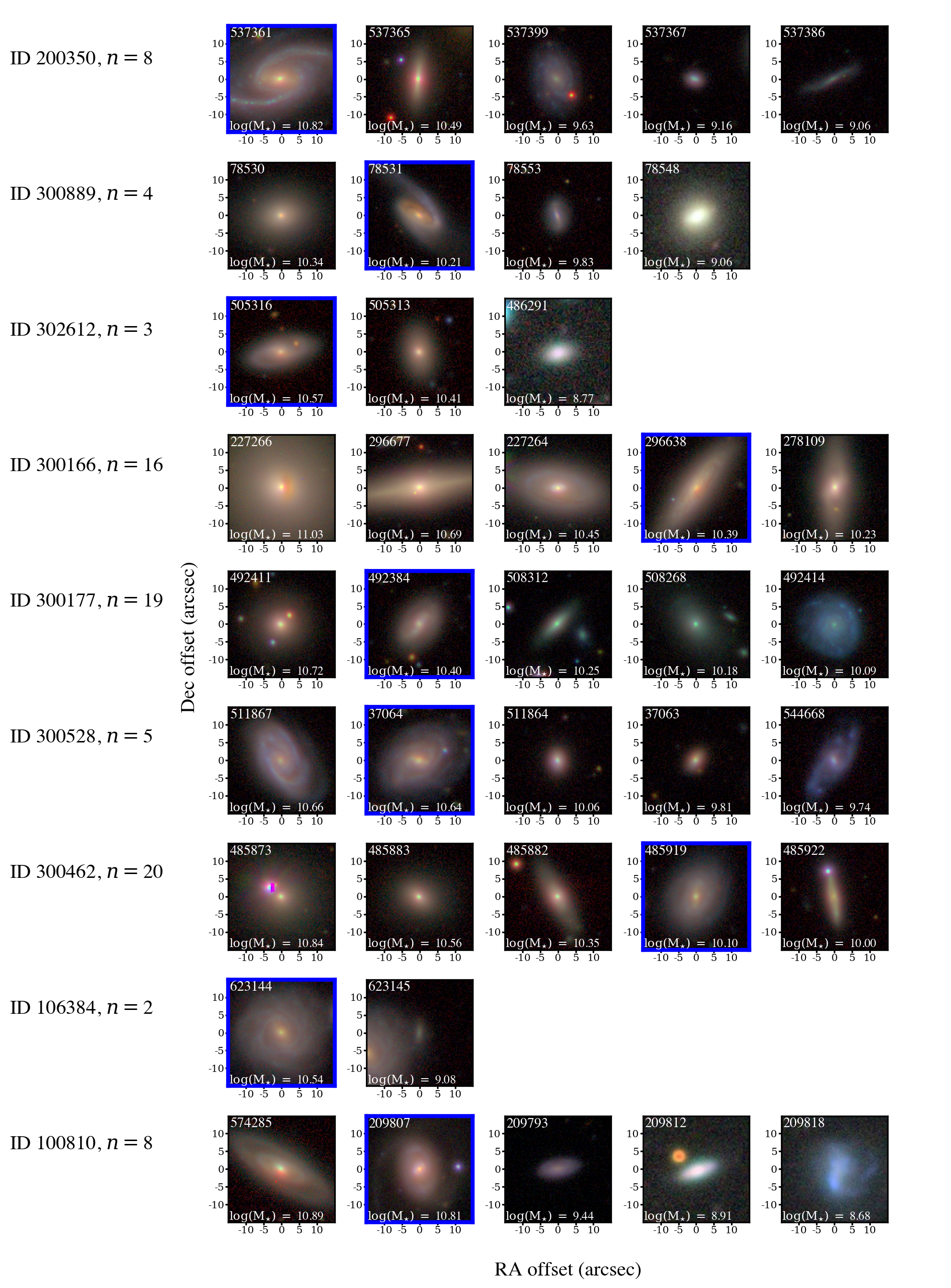}
    \caption{The images of nine GAMA galaxy groups where GAMA MWAs are located in \citep{Robotham2011} (except the isolated GAMA MWA 551655). In each galaxy group, the galaxies are sorted by stellar mass and up to five galaxies are shown. $n$ is the total number of galaxies being observed by GAMA in the galaxy group. Each image of a galaxy has a GAMA catalogue ID on the top left and its stellar mass on the bottom left. The GAMA MWAs are highlighted by blue frames.}
    \label{fig:companion}
\end{figure*}

Fig. \ref{fig:companion} illustrates the GAMA galaxy groups \citep{Robotham2011} where nine GAMA MWAs are located (with the exception of the isolated GAMA MWA 551655). In each galaxy group, we sort all galaxies by stellar mass and only show the top-5 most massive galaxies. The GAMA MWAs are marked by blue frames. The images have been obtained from the HSC \citep{Miyazaki2018} or the KiDS \citep{Kuijken2019} surveys.

\bsp	
\label{lastpage}
\end{document}